\documentclass[12pt,preprint]{aastex}
\usepackage{txfonts}
\usepackage{amssymb}

\shorttitle{Orbital Period and Outburst Luminosity of Transient
LMXB} \shortauthors{Wu et al.}

\begin{document}
\title{Orbital Period and Outburst Luminosity of Transient Low Mass X-ray Binaries}
\author{ Y. X. Wu\altaffilmark{1,~2}, W.
Yu\altaffilmark{1}, T. P. Li\altaffilmark{2,~3,~4}, T.J.
Maccarone\altaffilmark{5} and X. D. Li\altaffilmark{6} }
\altaffiltext{1}{Shanghai Astronomical Observatory, 80 Nandan Road,
Shanghai, 200030, China. E-mail: wenfei@shao.ac.cn}
\altaffiltext{2}{Department of Engineering Physics \& Center for
Astrophysics, Tsinghua University, Beijing, China.}
\altaffiltext{3}{Department of Physics \& Center for Astrophysics,
Tsinghua University, Beijing, China.} \altaffiltext{4}{Particle
Astrophysics Lab., Institute of High Energy Physics, Chinese Academy
of Sciences, Beijing, China.} \altaffiltext{5}{School of Physics and
Astronomy, University of Southampton, Southampton, Hampshire SO17
1BJ, UK.} \altaffiltext{6}{Department of Astronomy, Nanjing
University, Nanjing 210093, China.}

\begin{abstract}
In this paper we investigate the relation between the maximal
luminosity of X-ray outburst and the orbital period in transient low
mass X-ray binaries (or soft X-ray transients) observed by the {\it
Rossi X-ray Timing Explorer (RXTE)} in the past decade. We find that
the maximal luminosity (3--200~keV) in Eddington unit generally
increases with increasing orbital period, which does not show a luminosity saturation but in general agrees with theoretical prediction. The peak luminosities in ultra-compact binaries might be higher than those with orbital period of 2--4~h, but more data are needed to make the claim. We also find that there is no significant difference in the 3--200~keV outburst peak luminosity between
neutron star systems and black hole systems with orbital periods
above 4h; however, there might be significant difference at
smaller orbital period where only neutron star systems are observed and radiatively inefficient accretion flow is expected to work at the low luminosities for black hole accreters.

\end{abstract}


\section{Introduction}
Low mass X-ray binaries (LMXB) contain primary stars which are
either neutron stars (NSs) or black holes (BHs) and secondary stars
typically less than 1~M$_{\odot}$. The low mass companion stars are
usually main sequence or subgiant stars, and also include white
dwarfs or substellar mass objects. The secondary transfers matter to
the NS or BH primary via Roche lobe overflow followed by disk
accretion onto the primary. Soft X-ray transients (SXTs) are a
subset of LMXBs which spend most of their time in quiescence, and
occasionally exhibit outbursts during which their luminosities
increase by several orders of magnitude \citep[see reviews
by][]{TL95,vM95,MR06}. During outbursts, SXTs go through a variety
of ``canonical'' X-ray spectral states \citep[see
e.g.][]{Tan72,vdK95,FBG04,MR06,Don07} based on their spectral and
timing characteristics. The high/soft (HS) state corresponds to a
relatively high luminosity, the X-ray spectrum dominated by a soft thermal component and $<\sim5\%$ rms variability. The low/hard (LH)
state, on the other hand, is characterized by a relatively low
luminosity, the spectrum dominated by a hard non-thermal power-law component and $>\sim20\%$ rms variability. There are also
intermediate (IM) or very high (VH) states identified in the BH
systems, of which the X-ray spectrum is composed of both a steep
power-law component and a thermal component. SXTs usually enter the
HS state from the LH state during the rising phase of an outburst
and return to the LH state during the outburst decay. Past
observations have also shown that a source can remain in the LH
state throughout an outburst, which is usually of low luminosity
\citep{Bro06}.

The most popular model for SXT outburst is the disk thermal
instability (DTI) model \citep[see the review of][and references
therein]{Las01}. In this model, for a certain range of surface
density, the accretion disk can be either in a cool state where the
gas is neutral or in a hot state where the gas is ionized. In the
cool state the mass is accumulated in the disk until a surface
density threshold is reached, above which only the hot state is
possible. An outburst is then triggered, and the matter stored in
the disk is accreted by the compact object. Although the model is
generally adopted, diverse outburst properties are seen in the same
source during different outbursts, suggesting that the SXT outbursts
are complex.

If during an outburst the compact star accretes all of the material
stored in the accretion disk, the total output energy of the
outburst should be proportional to the mass in the accretion disk,
assuming that the radiative efficiencies of accretion process among
different outbursts remain the same. If the outburst profiles are
comparable and the peak luminosity is scaled to the total output
energy during outburst, we expect a correlation between the outburst
peak luminosity and the disk mass among outbursts. Previous X-ray
observations have shown that this is probably true. In the special
black hole transient GX~339$-$4, an empirical linear relation
between the peak flux of the LH state at the beginning of an
outburst and the outburst waiting time since the previous outburst,
and a correlation between the peak flux of the LH state during an
outburst rise and the outburst peak flux were found
\citep{YFK07,Wu09}. Actually the latter, namely the correlation
between the luminosity of LH-to-HS state transition and the peak
luminosity of the following HS state, has been confirmed for about
20 persistent and transient Galactic X-ray binaries in a luminosity
range spanning by 2 orders of magnitude \citep{YY09}. Combining the
two correlations, we expect a positive correlation between the
outburst peak luminosity and the mass stored in the disk responsible
for a certain outburst.

The maximal mass that can be stored in the disk before an outburst
is limited by the size of the Roch-lobe in these LMXBs. The size of
the Roch-lobe is reflected in the orbital period ($P_{\rm{orb}}$).
The binary separation is given by $P_{\rm{orb}}$ based on the
Kepler's law, and the Roche geometry is further determined by the
binary separation and the mass ratio of the primary and the
secondary \citep[see][]{Fra92}. In various models, the radial size
of the accretion disk has a close relation with $P_{\rm{orb}}$
\citep[see, e.g.][]{Pac77,Whi88,Lub91,Fra92}, and in general a
longer $P_{\rm{orb}}$ indicates a larger radius allowed for the
disk. The properties of the secondary for a given $P_{\rm{orb}}$ are
predicted by binary evolution theory. The mean density $\rho$ of the
secondary which fills its Roche lobe satisfies the approximate
relation of $\rho \cong 110P_{hr}^{-2}$ g cm$^{-3}$, in which
$P_{hr}$ is $P_{\rm{orb}}$ in the unit of hours \citep{Fra92}. The
mechanisms driving the mass transfer in the binaries are different
for different $P_{\rm{orb}}$, which could be the expansion of the
donor as it evolves away from the main sequence, or the loss of
orbital angular momentum through gravitational radiation or magnetic
braking \citep[e.g.,][]{Kin88,Kin96}. The amount of mass in the
accretion disk has been suspected to increase with increasing
$P_{\rm{orb}}$ and primary mass \citep[see, e.g.][]{MM00,PZ04}.

Theoretical calculation in \citet{Mey04} showed that the outburst
peak luminosity varies with $P_{\rm{orb}}$. \citet{PZ04} studied the
BH transients in \citet{Gar03} and showed that the peak luminosity
indeed varies with $P_{\rm{orb}}$, but not exactly as the theories
predicted. Similar study was also performed by \citet{Sha98} with
the peak luminosity taken from \citet{CSL97} (hereafter CSL97). It is worth noting that
the theoretical relations between $P_{\rm{orb}}$, the mass in the
disk and the outburst peak luminosity can only hold in an average
sense since SXT outburst varies from one to another, and the
Eddington limit on the outburst peak luminosities in both the LH
state and the HS state would cause a flat top of the outburst peak
luminosity because of luminosity saturation, although this probably
has not been seen \citep[e.g.,][]{YY09}.

CSL97 presented a systematic study of the light curves of
SXT outbursts before the {\it RXTE} era. Those transient outbursts
were observed by {\it Ariel~5}, {\it EXOSAT}, {\it Ginga}, {\it
CGRO} and so on from the 1960s to the middle 1990s. In the past
decade, X-ray observations of transient sources have been greatly
enriched. The missions such as {\it BeppoSAX}, {\it RXTE}, {\it
INTEGRAL} and {\it Swift} have discovered dozens of new transients.
Among them, {\it RXTE} has performed a lot of sensitive, pointed
observations of transient outbursts in the past 15 years. A
comprehensive study of these SXT outbursts observed with {\it RXTE}
is very necessary. In this paper, we present our systematic study of
the peak luminosity, the maximal HS state luminosity and the maximal
LH state luminosity (in Eddington unit) in relation to
$P_{\rm{orb}}$ for the transient sources listed in the LMXB
catalogue by \citet{Liu07} (hereafter LPH07).

\section{Data Analysis and Results}

\subsection{Source Selection }
Our source selection is based on the catalogue of LMXBs (LPH07). We
first selected sources identified as transients with known orbital
period $P_{\rm{orb}}$. We obtained a list of 40 sources, 20 of which
contain a BH or a BHC and the other 20 are thought to contain a NS.
This led to the initial sample, in which 22 sources were listed in
CSL97, including 17 confirmed SXTs and 5 possible SXTs.
Among the 5 possible SXTs, 3A~1516-569, known as Cir~X-1, probably
consists of a subgiant companion star with a mass of
3--5~M$_{\odot}$ or even higher \citep{Joh99,Jon07}, and its
periodic outbursts are often suggested to be the result of a highly
eccentric orbit, rather than disk ionization instability. We
excluded it from our source list. The other 4 sources, namely
MXB~1659-298, 4U~1659-487 (GX~339-4), 1A~1744-361, and GS~1826-238,
either have been taken as SXTs \citep[e.g.][]{Hom05,Bha06} or have
shown clear similarities to SXTs, so we included them. The remaining
18 sources in the list of 40 sources were not recorded in
CSL97. GRO~J1744-28 is excluded due to its characteristics
of both a pulsar and a type II burster
\citep[e.g.][]{Kou96,Gil96,Str96,Can96}. 4U~1755-33 and GRS~1747-312
are included in our sample because their transient natures are not
distinct from SXTs. The other 15 sources, including 7
accretion-powered millisecond pulsars (APMSPs), were discovered
after 1996 and are contained in our study. APMSP here refers to
those showing persistent (the first six APMSPs) or predominant
episodes (HETE~J1900.1-2455) of pulsations during outbursts, rather
than burst oscillations or the more recent intermittent pulsations
seen in Aql~X-1 and the NGC~6440 X-ray source \citep{Cas08,Alt08}.
SWIFT~J1756.9-2508 \citep{Kri07} is the eighth APMSP but not taken
into account due to its unknown distance. In total, our sample
includes 38 sources (20 BH or BHC systems and 18 NS systems) with
known $P_{\rm{orb}}$ (see Table~\ref{tbl1}). We have 3 more BHs and
BHCs than the list in \citet{MR06} (hereafter MR06).

In Table~\ref{tbl1} we list the X-ray flux (the peak flux or the
observed flux in the range between 2-10~keV, unless otherwise
indicated), the mass, the distance and the orbital period taken from
LPH07 for each of sources we selected.

\subsection{Flux Measurements from the {\it RXTE} PCA and HEXTE Energy Spectra}
As shown in the second column in Table~\ref{tbl1}, no outburst was
seen in some of the sources in the past decade. On the other hand,
two recently discovered sources CXOGC~J174540.0-290031 and
AX~J1745.6-2901 lack {\it RXTE} pointed observations. This leaves a
list of 25 sources whose outbursts have been observed with the {\it
RXTE} pointed observations, among which 9 are BH systems and 16 are
NS systems.

In order to estimate the outburst peak fluxes of these sources with
{\it RXTE} pointed observations, for each source we first determined
the date of the brightest outburst peak from all the PCA
observations in 2--9 keV from daily averaged PCA light curves. This
was checked with the {\it RXTE}/ASM light curve to guarantee that
the date identification is correct. For those sources discovered
before the {\it RXTE} era, the brightest outburst may be not the
ones observed by the {\it RXTE}. One example is GS~1354-64, which
showed a bright outburst entering the HS state in February, 1987 and
a dim LH state outburst in November, 1997. We only determined the
outburst peak flux with the {\it RXTE} observations. For some other
sources, the corresponding {\it RXTE} observations may not cover the
peak of the brightest outburst. For instance, the {\it RXTE}
observations of IGR~J00291+5934 and XTE~J2123-058 only covered the
outburst decays. In such cases, we took the first observation of
each source to estimate the outburst peak flux. In some NS systems
there were type I X-ray bursts. While the PCA daily averaged rates
include photon counts from type-I X-ray bursts, the contribution
from bursts is small. For the four sources in which bursts
contributed to the peak rates -- SAX~J1808.4-3658, XTE~J1814-338,
GS~1826-238 and HETE~J1900.1-2455 -- we checked the light curves to
make sure that we choose the observations consistent with the
maximal persistent PCA count rates. We then entered the date
corresponding to the observation of the outburst peak for each of
the 25 sources in Table~\ref{tbl2}.

The hardness ratios between the HEXTE (15-250~keV) and PCA (2-9~keV)
count rates were used to determine the spectral states as in
previous works \citep[e.g.][]{YD07}. The hardness ratios in the LH
state (on the order of 0.1) are ten times those of the HS states
(typically around 0.01). The VH state in BH systems has a hardness
ratio between those of the LH state and the HS state, associated
with a flux spike in the light curve. An outburst peak of BH
transients could be any of the three states, and that of NSs could
be either the LH state or the HS state. We identified the spectral
state corresponding to each outburst peak (Table~\ref{tbl2}). In
order to compare the peak fluxes of the same states in different
sources, we also analyzed the observations corresponding to the
maximal fluxes of different states if data permit. The maximal flux
of the LH state during the rise or the decay of an outburst is
marked as ``r'' or ``d'' in Table~\ref{tbl2}. Because the LH-to-HS
state transition during an outburst rise occurs usually at a higher
flux than the HS-to-LH state transition during the decay
\citep[known as ``hysteresis'',
e.g.][]{Miy95,Now95,Hom01,SHS02,MC03}, the HS-to-LH state transition
flux was taken as the lower limit of the maximal flux of the LH
state in case that the LH-to-HS state transition during the rise was
not covered by the {\it RXTE} pointed observations. We also
cross-checked with previous studies (e.g. \citealt{Par04} for
4U~1543-47; \citealt{Sob00} for XTE~J1550-564; \citealt{Mil02} for
XTE~J1650-500; \citealt{Bro06} and \citealt{Sha07} for GRO~J1655-40;
\citealt{Bel05} for GX~339-4; \citealt{YD07} for Aql~X-1) to make
sure that our identifications of spectral states were correct.

Next we analyzed the spectra of the selected observations to derive
the peak flux of the corresponding LH or HS state. The PCA/HEXTE
spectra provided as the {\it RXTE} standard products were fitted
with simple models. To ensure that the standard product provides the
representive energy spectrum, we checked the data to make sure that
the source flux is relative constant during the corresponding
observation and consistent with flux peak. The PCA spectra in
3--25~keV and HEXTE spectra above 20~keV were analyzed jointly, by
including a multiplicative factor to account for different
calibrations of the two instruments. A systematic error of 1\% was
added to the PCA/HEXTE data. The model consisting of a multicolor
blackbody (``diskbb'' in Xspec) plus a power-law was applied to BH
systems, and the model composed of a blackbody plus a power-law was
used for NS systems. A gaussian line with the energy bounded between
5.9 and 6.9~keV was sometimes included additionally to describe an
iron emission line component. In some cases the fits were
significantly improved by including a Fe absorption edge or a
smeared edge near 8~keV, or by replacing the power-law with a cutoff
power-law. The former was usually found in bright sources and the
latter in NS sources. Interstellar absorption was modeled using the
Wisconsin cross section (``wabs'' in Xspec). The hydrogen column
density in the direction of a certain source was fixed at the value
taken from the $N_{\rm{H}}$ tool provided by HEASARC website
(http://heasarc.gsfc.nasa.gov/cgi-bin/Tools/w3nh/w3nh.pl)
\citep{Kal05,DL90}. The best-fitting spectral parameters were listed
in Table~\ref{tbl2}. The total unabsorbed X-ray flux between
3--200~keV for each observation was derived (see Table~\ref{tbl3}).
Considering that the HEXTE spectrum in high energy band is dominated
by the contribution of background and might essentially add noise in
some cases, we also investigated if we cut off the HEXTE spectrum at
the energy where the source counts fall to less than $20\%$ of the
background level (typically at 30--40~keV), what the flux with the
best-fitting model extrapolated to 200~keV would be. We found the
difference is less than $0.5\%$, so we analyzed the energy spectra
in the 3--200~keV throughout to estimate the energy fluxes in the
3--200~keV range.

\subsection{Luminosity Estimates}
We calculated the X-ray luminosity in Eddington unit based on
$$\frac{L}{L_{\rm{Edd}}}=\frac{F\times 4\pi D^{2}}{1.3\times 10^{38}
M}$$ in which $F$ is the X-ray flux obtained in the spectral
analysis, $D$ is the source distance, $M$ is the mass of the compact
star in the unit of solar mass.

The uncertainties in the luminosity come from the uncertainties in
the mass, the source distance and the source flux, among which the
uncertainties in the mass and distance play a dominant role. BH mass
values and their uncertainties were mostly taken from LPH07, but for
XTE~J1118+480 and GRO~J1655-40, we also used the mass range from
MR06 in our study. The BH mass of XTE~J1859+226 was taken from MR06.
On the other hand, only the mass function are known for GS~1354-64
and 4U~1659-487. So we only had the lower limits on their BH masses.
The source distances listed in LPH07 and MR06 are generally
consistent. But for XTE~J1859+226 the distances are different, and
we used both in our study.

NS masses are not well measured in the transient LMXBs studied here.
Common understanding is that NSs have a canonical mass of about
1.4~M$_{\odot}$, but recent studies showed evidences for larger
neutron star masses in a few systems -- up to around 2~M$_{\odot}$
(e.g., $1.86\pm0.6$~M$_{\odot}$ for Vela~X-1, \citealt{Bar01};
$2.4\pm0.3$~M$_{\odot}$ for 4U~1700-37, \citealt{Cla02};
$>1.6$~M$_{\odot}$ for Aql~X-1, \citealt{Cor07}). Theory allows for
more than 2~M$_{\odot}$ of maximal NS mass under some reasonable
equations of state \citep[e.g., see the review of][and references
therein]{HP00}. We therefore assumed a NS mass range of 1.4 to
2.2~M$_{\odot}$ for all the NS LMXBs except XTE~J2123-058, which has
a better mass estimate. The distances of NS systems were taken from
LPH07.

\begin{figure}
\plotone{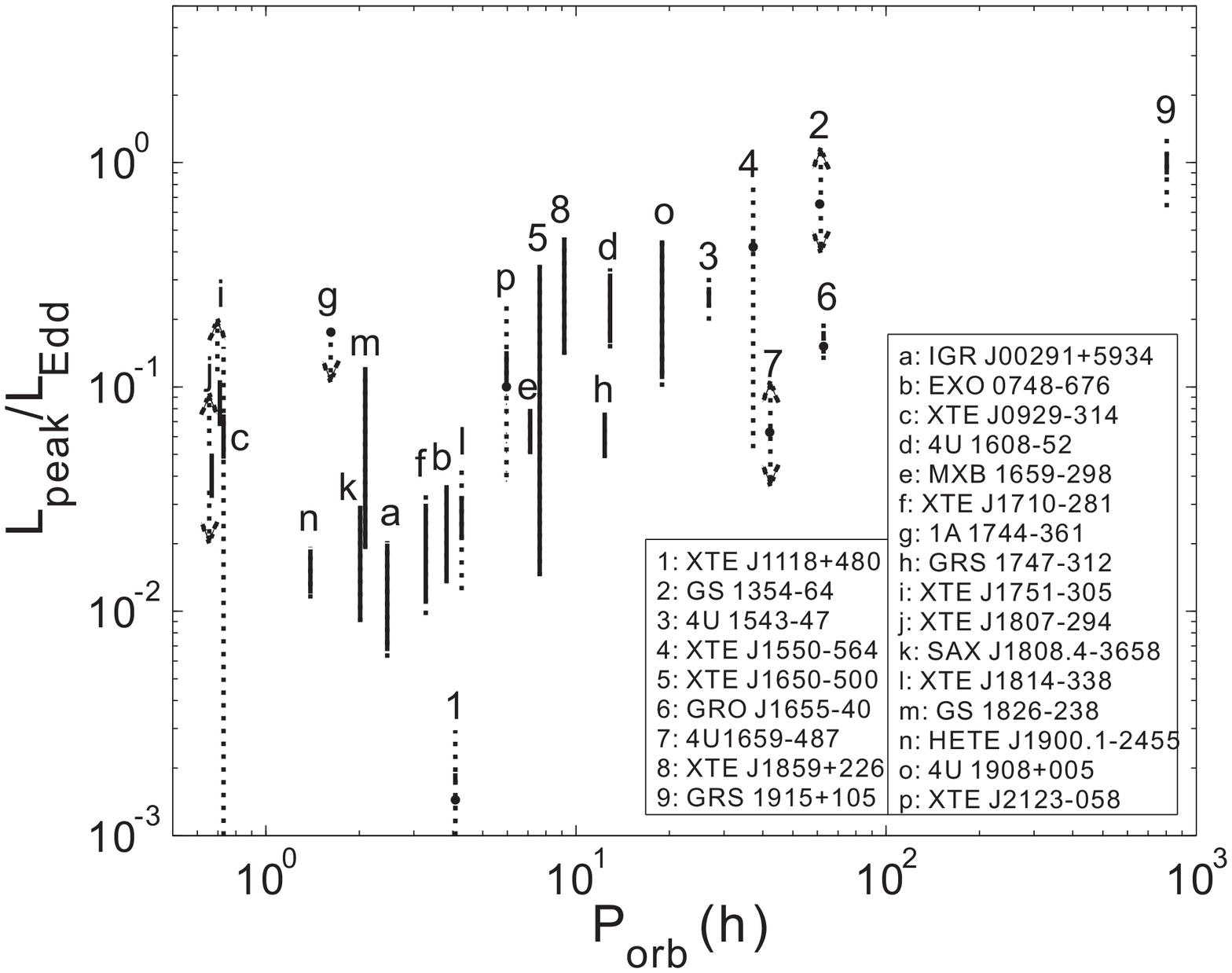} \caption{The peak luminosity
 (3--200~keV) as a function of the orbital period. The sources are
listed in Table~\ref{tbl3}. BHs and NSs are marked as numbers and
letters respectively, corresponding to the first column of
Table~\ref{tbl3}. Solid line shows the range of the source
luminosity and dotted line shows the uncertainty range. Arrow
represents the upper or lower limit. Double arrows indicate the
values are uncertain. \label{peaklum}}
\end{figure}

\begin{figure}
\plotone{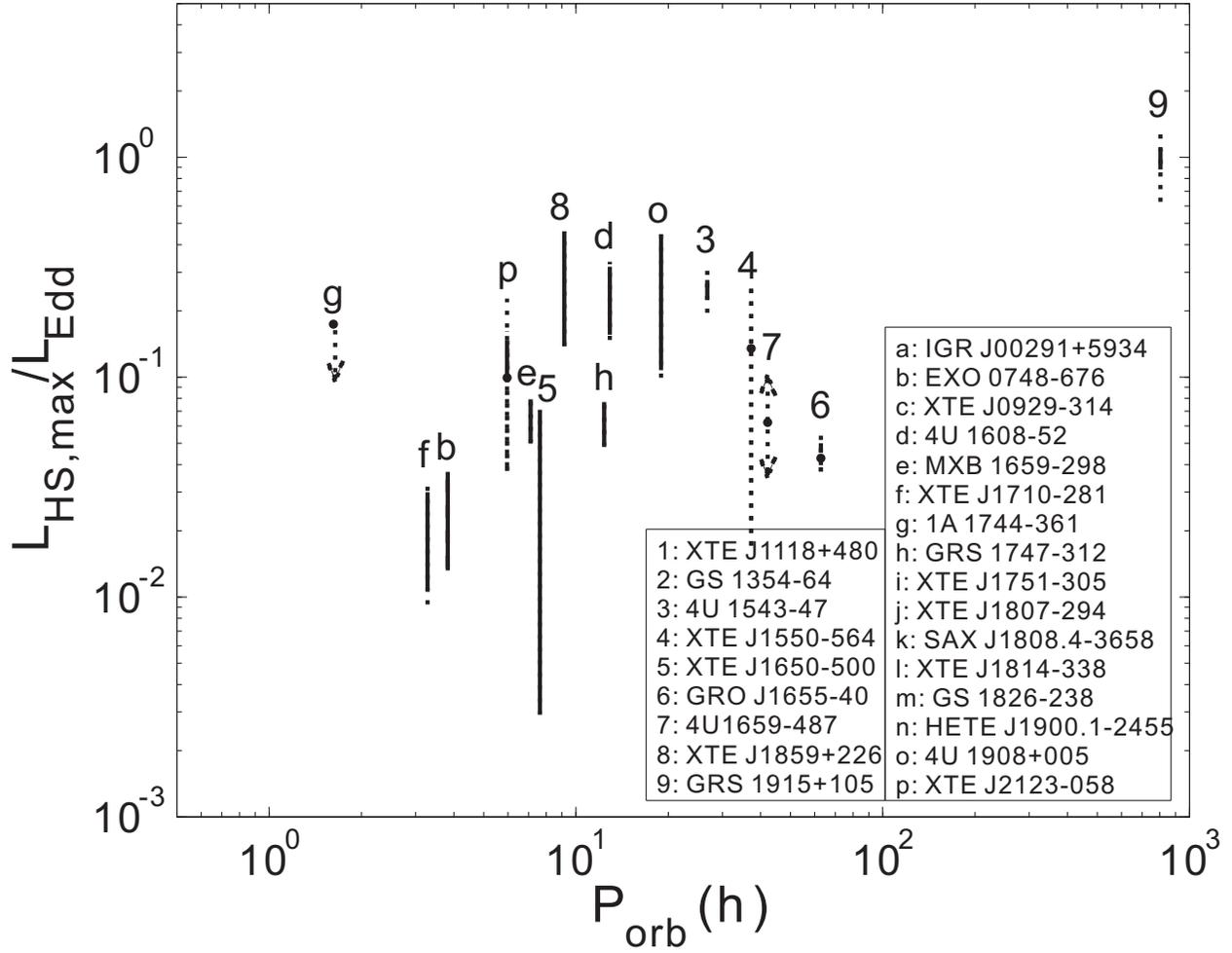} \caption{The maximal luminosity
 (3--200~keV) in the HS state as a function of the orbital period.
 Notation is the same as that in Figure~\ref{peaklum}. \label{hslum}}
\end{figure}

\begin{figure}
\plotone{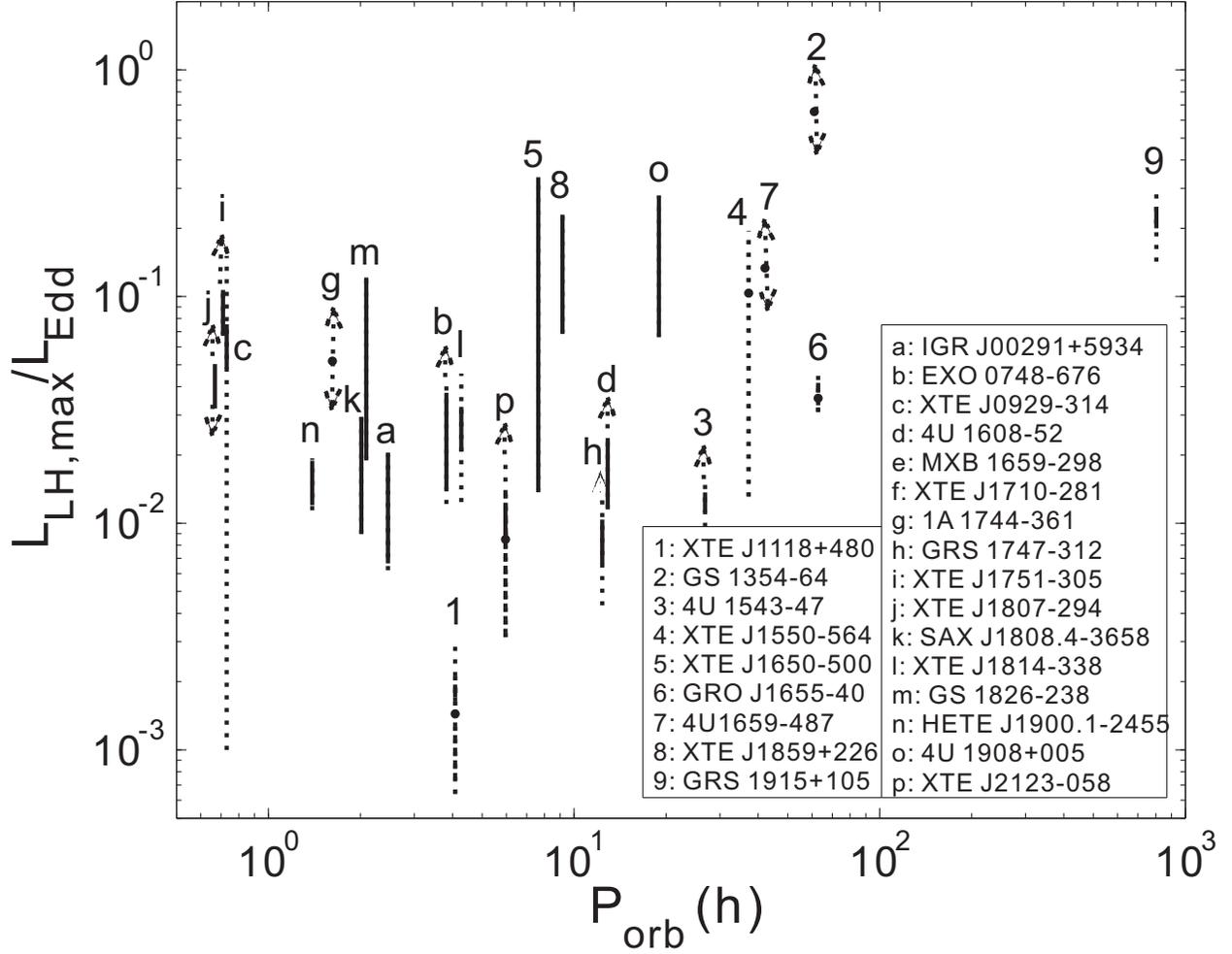} \caption{The maximal luminosity
 (3--200~keV) in the LH state as a function of the orbital period. Notation is the same as that in
 Figure~\ref{peaklum}. The luminosity of HS-to-LH state transition during the decay of an
outburst, is taken and plotted as the lower limit of the maximal LH
state luminosity if the LH state in the rise was not observed.
\label{lhlum}}
\end{figure}

We calculated source luminosities from the flux measurements. The
outburst peak luminosity, the maximal HS state luminosity and the maximal LH state luminosity in Eddington unit as a function of
$P_{\rm{orb}}$ are plotted in Figure~\ref{peaklum}--\ref{lhlum}, respectively. We
mark each source as a number for a BH or a letter for a NS, which
are shown in the first column of Table~\ref{tbl3} as well. For most
sources of which only a mass range or distance range is known, we
plot the corresponding luminosity range as a solid line and the
uncertainty range as a dotted line. The upper or lower limit on
luminosity is shown as an arrow. The results are plotted as double
arrows for GS~1354-64 (``2'') and 4U~1659-487 (``7''), because both
their masses and distances are given as lower limits and therefore
the luminosity estimates are not certain. For XTE~J1807-294 (``j''),
a distance of 8~kpc is assumed, hence it is marked with double
arrows. The luminosities of the above three sources are only shown
in the plot, but not used in our further analysis. It is worth
noting that the luminosity of the HS-to-LH state transition during
the decay of outburst was taken as the lower limit of the maximal LH
state luminosity due to the hysteresis effect.

We made a comparison between the peak luminosities derived here and those in the classical work of CSL97. Table~8 in CSL97 listed the value of $\log (L_{\rm p}/L_{\rm Edd})$ for 24 transient sources, whose outbursts were detected between 1967 and 1996. Our results are more accurate in at least two aspects: 1) the energy band studied in CSL97 was 0.4--10~keV and the observed fluxes obtained by different X-ray instruments were converted to this energy range. While in our work the flux is measured through  spectral modeling of the same instruments, with a broader energy coverage of 3--200~keV; 2) the uncertainties are estimated in our work, which are not present in CSL97. There are 6 sources covered by both samples. The maximal outburst peak luminosities of 4U~1543-47 and 4U~1608-52 in CSL97 are about twice as large as the corresponding values in our work. But in logarithmic scale such differences would not significantly affect the overall relation. For the other four sources, the peak luminosities we measured are larger than or comparable to those in CSL97.

\subsection{Outburst Peak Luminosity vs. Orbital Period}
For the peak luminosity of each source, the corresponding luminosity
range (hard boundary) and luminosity uncertainty range (soft
boundary) are shown as a solid line and a dotted line in
Figure~\ref{peaklum}, respectively. $L_{\rm{peak}}/L_{\rm{Edd}}$ on
the whole is higher for a source with a larger $P_{\rm{orb}}$, except the clear outlier XTE~J1118+480 with $P_{\rm{orb}}$ of $\simeq4.1$~h and $L_{\rm{peak}}$ of $\sim$0.001~$L_{\rm{Edd}}$. The highest $L_{\rm{peak}}/L_{\rm{Edd}}$ at the largest $P_{\rm{orb}}$
(GRS~1915+105, ``9'') is close to the Eddington limit. In order to further investigate the statistical properties of these samples, we attempted to fit the relation on the logarithmic scale. The data with only upper or lower limits were excluded, the median value of the hard boundary was
taken in the fitting, and the soft boundary was used to calculate the uncertainty. In the logarithmic axis the two-sided uncertainties would appear asymmetric and we used the larger one as the weight in the fitting.

\begin{figure}
\plotone{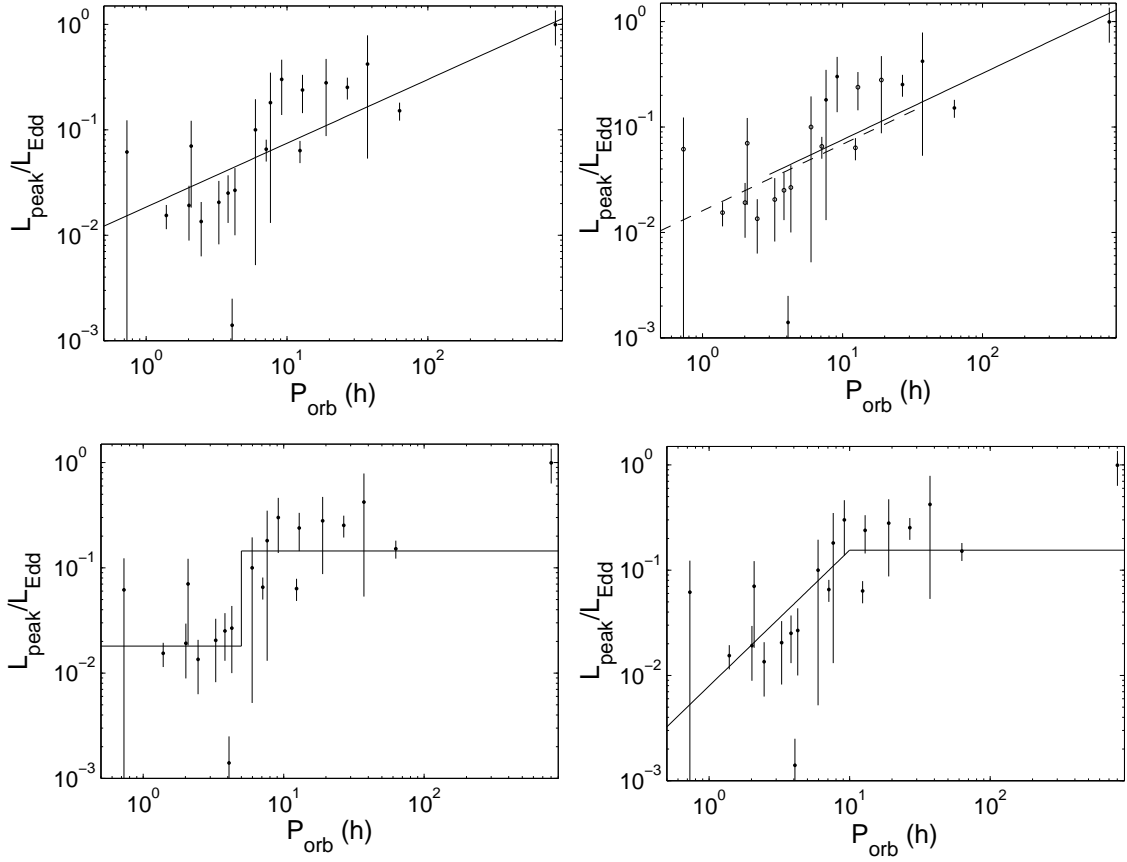} \caption{{\it Upper left panel:} the best-fitting linear model for
the overall sources in the log scale. {\it Upper right panel:} the lines fitted to
the BHs (shown as a solid line) and NSs (shown as a dashed line). {\it Lower left panel}: the best-fitting step function. {\it Lower right panel}: The best-fitting model consisting of a line and a constant, with the break period fixed at 10~h. The uncertain values in Figure~\ref{peaklum} (``2'',``7'',``g'',``i'',``j'') have been excluded from fitting for all the cases. In the latter three cases, the outlier at the orbital
period of 4.1h, XTE~J1118+480 (``1''), is excluded from fitting.
\label{fitlin}}
\end{figure}

First we fit with a straight line, which gives ${\rm
log}L_{\rm{peak}}/L_{\rm{Edd}}=(-1.80\pm0.11)+(0.64\pm0.08){\rm
log}P_{\rm{orb}}$ (see the upper left panel in Figure~\ref{fitlin}),
with a reduced $\chi^{2}=28.1/18=1.56$. The significance of the F-test is
$10^{-7}$, rejecting the null hypothesis that the slope is zero for
the overall sample. A positive correlation does exist for
$L_{\rm{peak}}/L_{\rm{Edd}}$ and $P_{\rm{orb}}$. We excluded the outlier XTE~J1118+480 and fit again. The fit gives ${\rm log}L_{\rm{peak}}/L_{\rm{Edd}}=(-1.78\pm0.11)+(0.63\pm0.08){\rm
log}P_{\rm{orb}}$, with a reduced $\chi^2=1.39$. The goodness of fit is improved but not significantly.

It appears that some sources in certain period regimes deviate from the linear relation. For example, not all the sources at small orbital periods show low luminosities. For the sources between 10~h and 100~h, the relation might be a ``flat top'' instead of a rise. Considering of these deviations, alternative models are also employed to fit the relation after excluding the outlier XTE~J1118+480. The plot indicates that $L_{\rm{peak}}/L_{\rm{Edd}}$ may sharply increase at $P_{\rm{orb}}\sim 5$~h. Therefore we tried the step function, i.e. a discontinuous fit with a smaller constant at short $P_{\rm{orb}}$ and a larger constant at long $P_{\rm{orb}}$. The best-fitting result is (shown in the lower left panel in Figure~\ref{fitlin})
$$\left \{ \begin{array}{ll}
L_{\rm peak}/L_{\rm Edd}=0.018 & \rm{if~{\it P}_{orb}<5~h} \\
L_{\rm peak}/L_{\rm Edd}=0.145 & \rm{if~{\it P}_{orb}\geq 5~h}
\end{array} \right.$$
and the reduced $\chi^{2}=2.81$. The step function fit the data worse than the straight line. Allowing the reduced $\chi^{2}$ increase by $\sim1$, we roughly estimated the uncertainties of the transition period as $5\pm2$~h.
We modified the model as a linear model saturating at a constant at long $P_{\rm{orb}}$. This model is used to describe the empirical relation of $\log L_{\rm{peak}}/L_{\rm{Edd}}$ and $\log P_{\rm{orb}}$ for BH X-ray binaries in \citet{PZ04}. The constant represents upper boundary for the transient luminosity. Under this model fit, if the break period is not constrained, the best-fitting value of this parameter turns out to be very large, making it essentially a pure linear model. If we fixed the break period as 10~h as in \citet{PZ04}, the best-fitting model can be expressed as  (shown in the lower right panel in Figure~\ref{fitlin})
$$\left \{ \begin{array}{ll}
\log L_{\rm{peak}}/L_{\rm{Edd}}=1.29\log P_{\rm{orb}}-2.10 & \rm{if~{\it P}_{orb}<10~h} \\
\log L_{\rm{peak}}/L_{\rm{Edd}}=-0.81 & \rm{if~{\it P}_{orb}\geq 10~h}
\end{array} \right.$$
and the reduced $\chi^{2}=2.34$. We found that the linear model describes the relation best among the three models, in spite of the plausible deviations. Our fits do not support that there exists a break orbital period, or a saturated luminosity. The longest orbital period systems can be pushed up to Eddington luminosity or perhaps slightly above the Eddington luminosity, depending on the bolometric corrections.

We tried to investigate whether there is a systematic luminosity
offset between BHs and NSs. We fixed the slope at 0.63, the
best-fitting value for the overall data, and fit the data of BHs and
NSs respectively. The results are shown in the upper right panel of
Figure~\ref{fitlin}. The intercepts are $-1.75\pm0.11$ for BHs and
$-1.79\pm0.05$ for NSs. They are consistent with being the same. We
noticed that in the regime where $P_{\rm{orb}}$ is less than 4h,
there are only NS sources, mostly APMSPs. We then fit the NS sources
with $P_{\rm{orb}}$ larger than 4~h and got $-1.75\pm0.08$. Again,
the intercepts of BHs and the NSs are consistent with the same. The
absence of BH systems in the small $P_{\rm{orb}}$ regime may
indicate that a BH system with a small $P_{\rm{orb}}$ tends to have
outbursts with peak fluxes below the sensitivity of the X-ray
monitoring missions in the past decade. This then implies that a
systematic luminosity offset between the BHs and the NSs at low
$P_{\rm{orb}}$ may exist, which is in the luminosity regime of below
$\sim 0.01$ $L_{\rm{Edd}}$.

We also studied the relation of the LH state peak luminosity
$L_{\rm{LH,max}}/L_{\rm{Edd}}$ vs. the orbital period
$P_{\rm{orb}}$, and the HS state peak
luminosity $L_{\rm{HS,max}}/L_{\rm{Edd}}$ vs. $P_{\rm{orb}}$. We
could not determine whether the peak luminosity of the LH state is
positively correlated with $P_{\rm{orb}}$, probably because we lack
the coverage of {\it RXTE} pointed observations.

\section{Discussion}
\subsection{Comparison with Theory}
It is not surprising that $L_{\rm{peak}}/L_{\rm{Edd}}$ tends to be
larger for systems with longer $P_{\rm{orb}}$. \citet{vP96} studied
disk instability in SXTs and concluded that an accreting NS or BH
system tends to be transient if the average mass transfer rate is
below a critical value which increases with $P_{\rm{orb}}$. This
implies that we tend to see transient sources with a larger
$P_{\rm{orb}}$ at a higher average mass accretion rate. For most of
the sources in our sample, the number of outbursts detected in the
past decade are about a few. Their recurrence times should not
differ too much and are roughly on the same order of magnitude. The
outburst peak luminosity is expected to correlate with the average
mass accretion rate for the sources we studied. Therefore we are
likely to see a positive correlation between
$L_{\rm{peak}}/L_{\rm{Edd}}$ and $P_{\rm{orb}}$ if the Eddington
limit is not reached.

\begin{figure}
\plotone{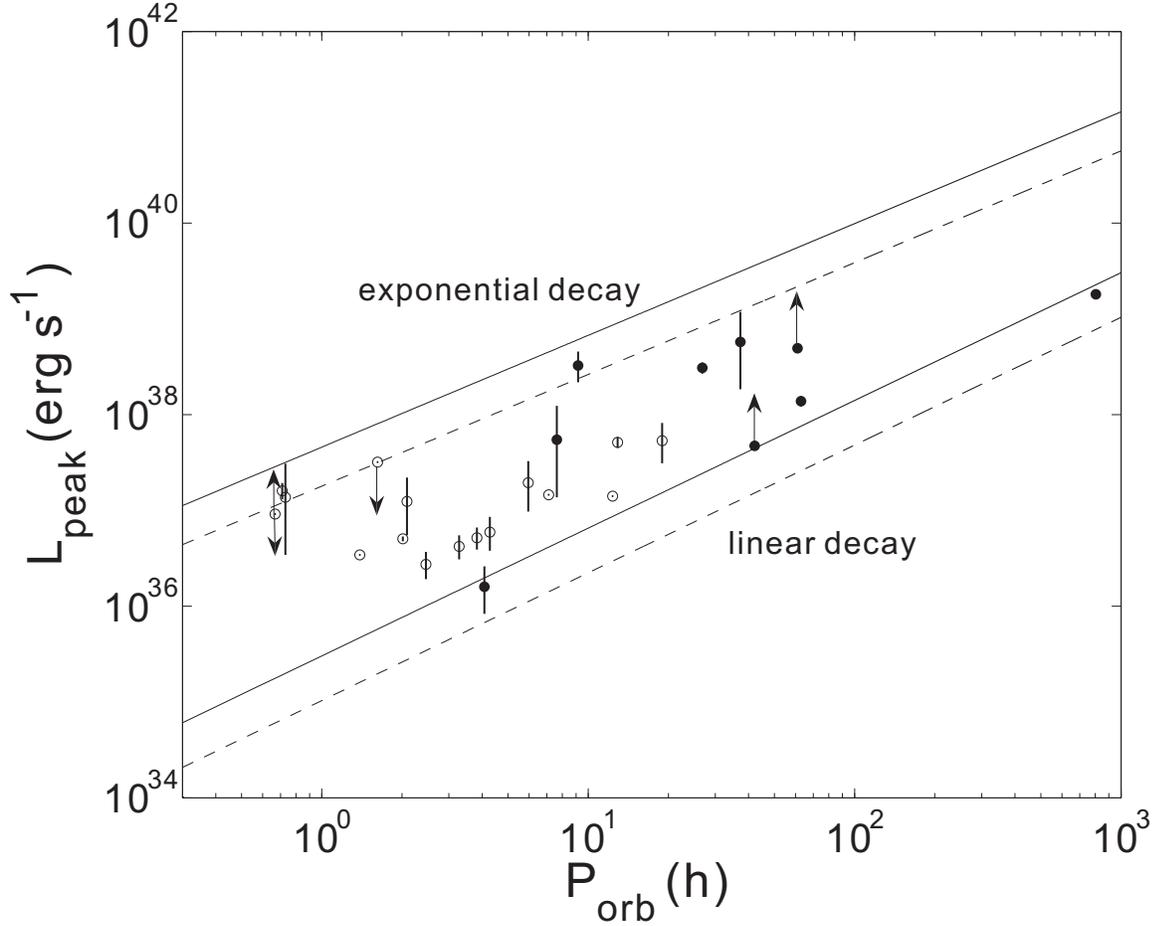} \caption{The theoretical outburst peak luminosity
of SXT as a function of orbital period. The upper pair of lines show
the peak luminosity corresponding to the exponential decay, the
lower pair corresponding to the linear decay. In each pair, the
solid and dashed lines show the peak luminosities for SXTs with
primary mass of 10~M$_{\odot}$ and 2~M$_{\odot}$, typical for BH and
NS systems respectively. The sources in Table~\ref{tbl3} are also
plotted, with dot representing BH and circle representing
NS.\label{theory}}

\end{figure}

The trend is also expected from the theoretical calculation of
\citet{KR98}. They showed that the light curves of SXTs can be
explained by disk instability if taking into account of irradiation
by the central X-ray source during the outburst. If the luminosity
is high, irradiation will be strong enough to ionize the entire disk
and the X-ray light curve would be roughly an exponential decay. If
the X-ray flux is too weak to ionize the whole disk, the outburst
light curve should be roughly a linear decay. Based on (31) and (32)
in \citet{KR98}, the peak luminosities for an outburst with an
exponential decay and linear decay can be calculated as
$$\left \{ \begin{array}{l}
L_{\rm{p,exp}}=2.7\times10^{38} R_{11}^{7/4}~\rm{erg~s^{-1}}\\
L_{\rm{p,lin}}=2.3\times10^{36} R_{11}^2~\rm{erg~s^{-1}}
\end{array} \right.
$$
respectively. Here $R_{11}$ is the maximal ionized disk radius in
units of $10^{11}$~cm, which is roughly the maximum of the outer
disk radius, usually taken as the tidal radius of the disk and
estimated as $80\%$ of the primary's Roche lobe radius.
According to \citet{Egg83}, the primary's Roche lobe radius can be calculated by the approximate analytic formula
$$R_1=\frac{0.49aq^{-2/3}}{0.6q^{-2/3}+\ln(1+q^{-1/3})},$$
in which $q$ is the mass ratio between the secondary and the primary $m_2/m_1$, and $a$ is the binary separation which can be conveniently expressed in the form
$$a=3.5\times 10^{10} m_1^{1/3} (1+q)^{1/3} P_{\rm orb}^{2/3}~{\rm cm}.$$
Assuming $q\sim0.1$ and combining the above equations, we get
$$\left \{ \begin{array}{l}
L_{\rm{p,exp}}=1.18\times 10^{37} m_1^{7/12} P_{\rm orb}^{7/6}~\rm{erg~s^{-1}}\\
L_{\rm{p,lin}}=6.42\times 10^{34} m_1^{2/3} P_{\rm orb}^{4/3}~\rm{erg~s^{-1}},
\end{array} \right.$$
where $m_1$ is the primary mass in units of solar mass, and $P_{\rm orb}$ is the orbital period in units of hour. We plot the
theoretical results as solid lines in Figure~\ref{theory} together
with the peak luminosities we measured. No source is observed beyond
the maximal peak luminosity. At short $P_{\rm{orb}}$,
$L_{\rm{peak}}$ tends to be close to the theoretical peak luminosity
corresponding to an outburst with an exponential decay, indicating
that the disks in these systems are likely entirely irradiated.
Systems with longer $P_{\rm{orb}}$ have $L_{\rm{peak}}$ close to the
theoretical peak luminosity for a linear decay, probably because the
disk in these systems are large and only the inner parts of the
disks are irradiated. This is consistent with the prediction of
\citet{KR98}.

\subsection{XTE~J1118+480}
Orbital period and peak luminosity in Eddington unit generally follow
a positive correlation, whereas XTE~J1118+480 is a outlier. Its 3--200~keV peak luminosity is only about $10^{-3}L_{\rm{Edd}}$, one order of magnitude lower than the other sources with similar $P_{\rm{orb}}$. The source is the BH SXT known with the shortest $P_{\rm{orb}}$ and almost the shortest distance, which remained in the LH state throughout the outburst we observed. During the outburst, it showed strong non-thermal radiation in the radio to optical band and a very low X-ray to optical flux ratio
\citep[e.g.][]{Gar00}. Another even fainter outburst was observed in
January, 2005 \citep{Rem05}, which proves that such faint outburst
is not occasional but typical for XTE~J1118+480. Several physical
interpretations have been proposed for this source, involving the
radiative inefficient accretion flow \citep{Esin01}, jet synchrotron
radiation in the LH state \citep{Mar01} and jet-disk coupling
\citep{Mal04}. A popular idea is that the bolometric correction to
its luminosity would be much larger than those for other sources
which stayed in the LH state throughout an outburst. However,
$L_{\rm{peak}}/L_{\rm{Edd}}$ of XTE~J1118+480 is rather low even
compared with the brightest LH state luminosities of other sources
(Figure~\ref{lhlum}), which suggests alternative interpretations.
One possibility is that the radiative efficiency for systems far
from the critical luminosity of state transitions is lower than that
for the brightest LH state systems --- the advection dominated
accretion flow model, for example, gives $L_X \propto \dot{m}^2$
\citep[e.g.][]{Esin97}. If this model applies to XTE J1118+480, then
the radiative efficiency of this system might be a factor of $\sim5$
smaller than the other black hole systems, putting its mass
accretion rate much closer to the other systems than its luminosity.

\subsection{Ultra-Compact Binary with Higher Luminosity?}
In Figure~\ref{peaklum} the complex phenomena below 4~h can be noticed: the peak luminosities for sources with orbital period between 2--4~h ---namely IGR~J00291+5934 (``a''), EXO~0748-676 (``b''), XTE~J1710-281 (``f'') and XTE~J1814-338 (``l'')---drop suddenly; while among the three sources with $P_{\rm{orb}}<1$~h, XTE~J1751-305 (``i'') has a lower limit of $L_{\rm{peak}}/L_{\rm{Edd}}$ which is one order of magnitude larger than those between 2--4~h, and the other two might have higher $L_{\rm{peak}}/L_{\rm{Edd}}$ but accurate distance measurements are not available. It is an interesting question whether the ultra-compact X-ray binary with orbital period smaller than 1~h can reach higher peak luminosity than the source in the period range of 2--4~hours.

Only according to this work, the evidences are not conclusive due to the below considerations. In the first place the number of sources is not enough in statistics, and their masses and distances are quite uncertain. For example the distance of XTE~J0929-314 (``c''), indicated as $10\pm 5$~kpc in LPH07, seems to lack of solid observational basis. Secondly for very close binaries the donor stars are likely to be hydrogen depleted, so that their Eddington luminosities are different from those in wider systems---in which accreting matter is hydrogen---by a factor of a few.
As a result $L_{\rm{peak}}/L_{\rm{Edd}}$ plotted here are not the actual values for the ultra-compact binaries. Thirdly we notice that all the three unltra-compact binaries with orbital period lower than 1~h are APMSPs. There may be a selection effect at short orbital period regime since the short period were determined from X-ray pulse arrival times\citep[e.g. ][]{Mar02}. At the same time these X-ray pulsars are very likely with non-isotropic emission from the polar region in addition to disk emission, and their apparent luminosities to the observer have a systematic increment. Lastly the low outburst peak luminosities observed in sources with orbital periods of several hours might be accidental. For example, IGR~J00291+5934 (``a'') has a radio counterpart \citep{Fen04} that indicates an outflow, which is likely radiatively inefficient. In addition, it might be at a larger distance than we thought \citep[7--10~kpc,see][]{Bur06}. Both factors would cause the low luminosity -- the former by reducing the intrinsic luminosity and the latter by underestimating the conversion factor from flux to lumniosity. Another example is EXO~0748-676 (``b''). It has been observed with
X-ray dips \citep{Par86}; strong inclination effects may cause a
lower luminosity as well. All the above factors would cause the ultra-compact binaries appear brighter than we expected from the empirical linear relation. Just based on our work it might be premature to claim that $L_{\rm{peak}}/L_{\rm{Edd}}$ in unltra compact binary system is essentially higher than that in the system not so compact. However it is a problem worthy of further investigation, because additional accretion stream component or slightly different accretion geometry might exist in ultra-compact X-ray binaries \citep[e.g. ][]{Zhang06}.

\subsection{BH and NS systems}
An interesting investigation is a comparison of BH and NS systems.
\citet{Men99} and \citet{Gar01} studied the quiescent luminosities
of NS and BH SXTs as a function of $P_{\rm{orb}}$, and found
systematically higher quiescent luminosities for NS SXTs. This is
consistent with the idea that the former has a hard surface and the
later has an event horizon instead. However, further studies show
that at least some of the NS SXTs can be as faint as BH SXTs in
quiescence \citep[e.g.][]{Jon07b}. In our study, we found no
significant difference in the outburst peak luminosities between BHs
and NSs systems at similar $P_{\rm{orb}}$. This indicates that the
radiative efficiencies of the accretion flows around NSs and BHs
during the outburst peaks are comparable, probably except for the
case of XTE~J1118+480 discussed above.

It is worth noting that all the BH transients we studied have an
orbital period larger than $\sim$ 4h. Significant difference in the
radiation efficiencies between BH systems and NS systems at low
luminosities and at small orbital periods can not be ruled out. BH
transient systems with $P_{\rm{orb}}\lesssim 4$h should have very
faint outbursts. Because for short $P_{\rm{orb}}$ SXTs, they should
have low mass transfer rate in order to remain as transients
\citep{vP96}, about $1\%\dot{M}_{\rm{Edd}}$ or below for BHs towards
small $P_{\rm{orb}}$ ($\lesssim 4$h). At such a low mass accretion
rate, the accretion flow would be radiatively inefficient and the
observed luminosity is far less than $1\% L_{\rm{Edd}}$. Therefore the
BH systems are expected dimmer than the NS systems at short
$P_{\rm{orb}}$. XTE~J1118+480, which is the faintest
($\sim0.1\%L_{\rm{Edd}}$, much smaller than other BHs) as well as the
shortest-period BH in our sample, shows some evidence of radiatively
inefficient accretion. This implies that few BH systems with an orbital period below 4h are observed may be the result of radiatively inefficient
flow at low mass accretion rate in BH systems.

\subsection{The Accretion Disk Mass}
Evidence for the connection between the mass in the accretion disk
when an outburst occurs and the peak luminosity of an outburst can
be inferred from the observations of black hole transient GX~339-4
\citep{YFK07,Wu09}. For the same compact star mass and companion
mass in a LMXB, $P_{\rm{orb}}$ indicates the size of the Roche lobe.
The mass stored in the accretion disk during quiescence is then
limited by the size of the Roche lobe. We show that there is an
overall positive correlation between $L_{\rm{peak}}/L_{\rm{Edd}}$
and $P_{\rm{orb}}$ in transient LMXBs. This supports the idea that
there is a correlation between the outburst peak luminosity and the
initial mass in the accretion disk \citep{YKF04,YFK07,YY09}. A
positive correlation between $L_{\rm{peak}}/L_{\rm{Edd}}$ and
$P_{\rm{orb}}$ therefore strengthens that the mass in the accretion
disk affects the outburst properties. Indeed, theoretical studies of
SXT outbursts have shown that a positive correlation between
$L_{\rm{peak}}/L_{\rm{Edd}}$ and $P_{\rm{orb}}$ exist for certain
model parameters \citep{Mey04}. Taking into account that there is a
correlation between the LH-to-HS transition luminosity, the
rate-of-change of luminosity, and outburst peak luminosity in X-ray
binaries \citep{YY09}, the mass in the accretion disk seems the most
important parameter that sets up the accretion processes during
outbursts.

As shown above, the relation between the allowed maximal outer disk radius $R_{\rm max,disk}$ and $P_{\rm orb}$ can be approximated as $R_{\rm max,disk}\sim P_{\rm{orb}}^{2/3}$. The best-fitting linear model in the logarithmic axis gives $L_{\rm peak}/L_{\rm Edd}\sim P_{\rm orb}^{0.64}$. Combining them we obtained $L_{\rm peak}/L_{\rm Edd}\sim R_{\rm max,disk}$, which indicate a nearly linear relation between the Eddington-scaled maximal peak luminosity and the allowed maximal outer disk radius. This is an interesting result and might provide constraints for some certain theoretical models and physical parameters. If total energy fluence of outburst is measured and total mass accreted to the compact star is estimated, the threshold of disk surface density for triggering outburst can be constrained.

\section{Summary}
We studied the relation between the outburst peak luminosity and the
orbital period in transient LMXBs. We analyzed the PCA/HEXTE spectra
of 25 LMXBs whose outburst peaks had been covered by {\it RXTE}
pointed observations, and measured the peak flux of the brightest
outburst of each source in the energy range between 3--200~keV. We
then estimated the luminosities in Eddington unit as well as their
uncertainties. The maximal luminosity of the LH and HS state were
also studied as well.

We found that $L_{\rm{peak}}/L_{\rm{Edd}}$ on the whole is
higher for sources with larger $P_{\rm{orb}}$, showing no sign of
luminosity saturation expected when approaching the Eddington limit
towards larger $P_{\rm{orb}}$. We fitted the relation in the logarithmic scale with three models: a  straight line model, a step function model and a straight line connected with a constant model. The linear relation can well describe the data.

We find the theoretic work by \citet{KR98} is quantitatively consistent with the relation of peak luminosity and orbital period we obtained. The contrast at short orbital period regime might indicate that the ultra-compact binaries have unexpected higher luminosities which, however, is not conclusive only based on our study due to a number of observational uncertainties. We have also compared $L_{\rm{peak}}/L_{\rm{Edd}}$ of NS and BH SXTs, and found that generally there is no significant difference between them. Therefore radiative efficiencies of the accretion flows are comparable for NS and BH with similar $P_{\rm{orb}}$ during outburst peaks. Since we have not seen BH systems with $P_{\rm{orb}}$ less than 4h, it is possible that BH transients with short $P_{\rm{orb}}$ have lower outburst peak luminosities than the NS systems, because the accretion flow might be radiative inefficient at low mass accretion rate.

\acknowledgments We would like to thank the RXTE Guest Observer
Facilities at NASA Goddard Space Flight Center (GSFC) for providing
the RXTE standard products and the ASM monitoring results. WY would
like to thank Jean Swank and Craig Markwardt of GSFC for useful
discussions and kind assistance. YW would like to thank Roberto
Soria, Piergiorgio Casella, Tomaso Belloni, Shuang Nan Zhang and Hua
Feng. This work was supported in part by the National Natural
Science Foundation of China (including 10773023, 10833002), the One
Hundred Talents project of the Chinese Academy of Sciences, the
Shanghai Pujiang Program (08PJ14111), the National Basic Research
Program of China (2009CB824800), and the starting funds of the
Shanghai Astronomical Observatory. The study has made use of data
obtained through the High Energy Astrophysics Science Archive
Research Center Online Service, provided by the NASA/Goddard Space
Flight Center.

\clearpage

\clearpage

\clearpage

\clearpage

\clearpage

\clearpage

\begin{deluxetable}{ccccccc}
\tablecaption{List of transient LMXBs with known orbital period
\label{tbl1}} \tablewidth{0pt} \setlength{\tabcolsep}{0.04in}
\tabletypesize{\scriptsize} \tablehead{ \colhead{Source} &
\colhead{Type} & \colhead{Year} & \colhead{$F\rm{_x}(\mu
Jy)$(2-10~keV)} & \colhead{$M$(M$_{\odot}$)} & \colhead{$D$(kpc)} &
\colhead{$P\rm{_{orb}}$(hr)} }

\startdata
GRO~J0422+32 & BH & 1992 & 2800 (8-13~keV) & 3.97$\pm$0.95 & 2.49 & 5.092\\
A~0620-00 & BH & 1975 & 50000 & 11.0$\pm$1.9 & 1.2 & 7.75\\
GRS~1009-45 & BH & 1993 & 800 (1-10~keV) & $>$3.9 & 1.5-4.5/5.7 & 6.84\\
XTE~J1118+480 & BH & 2000,2005 & 40 (2-12~keV) & 8.53 & 1.8 & 4.08\\
GS~1124-684 & BH & 1991 & 3000 & 4-11\tablenotemark{a} & 5.9 & 10.38\\
GS~1354-64 & BH\tablenotemark{b} & 1987,1997 & 5-120 & 5.75(f) & $\ge$27 & 61.07\\
4U~1543-47 & BH & 1971,1983,1992, & $<$1-15000 & 8.4-10.4 & 7.5 & 26.8\\
&& 2002 & & &  & \\
XTE~J1550-564 & BH & 1998,2000,2001, & 600-7000 & 10.5$\pm$1.0
& 5.3 & 37.25\\
&& 2002,2003 & & &  & \\
XTE~J1650-500 & BH\tablenotemark{c} & 2001 & 455 (0.5-10~keV) & $<$7.3/~4 & 2-6/2.6 & 7.63\\
GRO~J1655-40 & BH & 1994,1996,2005 & 1600 & 7.0/6.3$\pm$0.5 & 3.2/$<$1.7 & 62.88\\
4U~1659-487(GX~339-4) & BH & 1972-2007 &
1.5-900 & 5.8$\pm$0.5(f) & $>$6 & 42.14\\
4U~1705-250 & BH & 1977 & $<$2-3600 & 4.86(f) & 8.6 & 12.54\\
GRO~J1719-24 & BH\tablenotemark{d} & 1993,1995 & 1500 (20-100~keV) &
4.9 & 2.4 & 14.7\\
1A~1742-289 & BHC & 1975 & $<$9-2000 & \nodata & 10 & 8.356\\
CXOGC~J174540.0-290031 & BHC\tablenotemark{e} & 2005 & 0.2(2-8~keV)
& \nodata & \nodata & 7.9\\
4U~1755-33 & BHC & 1974-1996 & 100 & \nodata & 4-9 & 4.4\\
XTE~J1859+226 & BH & 1999 & 1300 & 7.4$\pm$1.1(f) & 7.6 & 9.16\\
GRS~1915+105 & BH & 1992- & 300 & 14$\pm$4 & 11.2-12.5 & 804\\
GS~2000+25 & BH & 1988 & $<$0.5-11000 & 4.8-14.4 & 2.7 & 8.26\\
GS~2023+338 & BH & 1989 & 0.03-20000 & 12$\pm$2 & 3.5 & 155.4\\
\hline IGR~J00291+5934 & NS (MP) & 2004 & 52 (2.5-25~keV) & \nodata
& 2.6-3.6 & 2.46\\
EXO~0748-676 & NS(B) & 1985- & 0.1-60 & \nodata & 5.9-7.7 & 3.82\\
XTE~J0929-314 & NS(MP) & 2002 & 36 & \nodata & 10$\pm$5 & 0.73\\
4U~1456-32(Cen~X-4) & NS(B) & 1969,1979 & 0.1-20000 &
1.3$\pm$0.6/ & 1.2 & 15.10\\
&&&&1.5$\pm$1.0&&\\
4U~1608-52 & NS(B) & 1976- & $<$1-110 & \nodata & 4.1/3.6 & 12.89\\
MXB~1659-298 & NS(B) & 1976,1999 & $<$5-80 & \nodata & 10 & 7.11\\
XTE~J1710-281 & NS(B) & 1998 & 2 & \nodata & 12-16 & 3.28\\
1A~1744-361 & NS(B) & 1976,1989,2003, & $<$25-275 & \nodata
& $<$9 & 1.62\\
&& 2004,2005 & & &  & \\
AX~J1745.6-2901 & NS(B) & 1996 & 0.4-2 (3-10~keV) & \nodata &
\nodata & 8.4\\
GRS~1747-312 & NS(B) & 1990,1996-1999, & 1.5-20 & \nodata & 9.5
& 12.36\\
&& 2004 & & &  & \\
XTE~J1751-305 & NS(MP) & 2002 & 60 (2-10~keV) & \nodata &
$\sim$8.5/$>$7 & 0.71\\
XTE~J1807-294 & NS(MP) & 2003 & 58 (2-10~keV) & \nodata & \nodata &
0.668\\
SAX~J1808.4-3658 & NS(MP,B) & 1998,2000,2002, & 110 & $<$2.27 &
2.5/3.4-3.6 & 2.014167\\
&& 2005,2008 & & &  & \\
XTE~J1814-338 & NS(MP,B) & 2003 & 13 & \nodata & 8.0$\pm$1.6 & 4.27462\\
GS~1826-238 & NS(B) & 1988- & 30 & \nodata & 4-8 & 2.088\\
HETE~J1900.1-2455 & NS(MP,B) & 2005 & 55 (2-20~keV)& \nodata & 5 & 1.39\\
4U~1908+005 (Aql~X-1) & NS(B) & 1978- & $<$0.1-1300 & \nodata & 5 &
18.95\\
XTE~J2123-058 & NS(B) & 1998 & 110 & 1.5$\pm$0.3/ &
8.5$\pm$2.5& 5.96\\
&&&&1.04-1.56&&\\
\enddata
\tablecomments{`Source'--the source name taken from LPH07;
`Type'--for neutron star, `B' represents X-ray burst source, `MP'
represents millisecond pulsar; `Year'--the years of outbursts or the
year of discovery; `$F\rm{_x}$'--the maximal X-ray flux, or the
range of observed X-ray fluxes (2-10~keV, unless otherwise
indicated), $1\mu
Jy=10^{-29}~\rm{erg~cm^{-2}~s^{-1}~Hz^{-1}}=2.4\times
10^{-12}~\rm{erg~ cm^{-2}~s^{-1}~keV^{-1}} $; `$M$'--the mass of the
central compact star, the letter `f' in the parentheses indicates
the mass function; `$D$'--the distance; `$P\rm{_{orb}}$'--the orbit
period. The data in the above four columns are from LPH07.}
\tablenotetext{a}{From \citet{Sha97} which is taken as reference for
BH mass in LPH07} \tablenotetext{b}{Classified as BHC in MR06, but
confirmed as BH in \citet{Cas04}.} \tablenotetext{c}{Classified as
BHC in MR06, but confirmed as BH in \citet{Oro04}.}
\tablenotetext{d}{Classified as BHC in MR06, but taken as BH in
LPH07 with a mass of 4.9~M$_{\odot}$ \citep{Mas96}. }
\tablenotetext{e}{Not listed in MR06, and discovered by
\citet{Mun05}. }
\end{deluxetable}
\clearpage

\begin{deluxetable}{cccccccccccccc}
\tablecaption{Spectral Parameters of transient LMXBs \label{tbl2}}
\tablewidth{0pt} \tabletypesize{\scriptsize}  \rotate
\setlength{\tabcolsep}{0.02in}
\renewcommand\arraystretch{1.5} \tablehead{ \colhead{Source} &
\colhead{State} & \colhead{MJD} & \colhead{ObsID} &
\colhead{$N_{\rm{H}}$} & \colhead{$T_{\rm{B}}$} &
\colhead{$N_{\rm{B}}$} & \colhead{$\Gamma_{\rm{PL}}$} &
\colhead{$N_{\rm{PL}}$} & \colhead{$E_{\rm{Fe~line}}$} &
\colhead{$\sigma_{\rm{Fe~line}}$} & \colhead{$N_{\rm{Fe~line}}$} &
\colhead{$\chi_{\nu}^2$/dof} & \colhead{additional details} }

\startdata

XTE~J1118+480 & P,LH & 51659 & 50407-01-03-01 & 0.013 & \nodata &
\nodata & 1.748$^{+0.010}_{-0.010}$ & 0.298$^{+0.006}_{-0.007}$ &
\nodata & \nodata & \nodata & 0.80/99 &
\nodata\\
GS~1354-64 & P,LH & 50787 & 20431-01-04-00 & 0.7 &
1.82$^{+0.07}_{-0.07}$ & 2.39$^{+0.27}_{-0.25}$ &
1.12$^{+0.06}_{-0.06}$ & 0.086$^{+0.014}_{-0.013}$ & \nodata &
\nodata & \nodata & 0.95/93 & PL cutoff 60~keV\\
4U 1543-47 & P,HS & 52445 & 70133-01-04-00 & 0.4 &
0.979$^{+0.005}_{-0.004}$ & 7876$^{+255}_{-132}$ &
2.63$^{+0.02}_{-0.03}$ & 5.60$^{+0.20}_{-0.39}$ &
6.00$^{+0.19}_{-0.34}$ & 0.83$^{+0.04}_{-0.15}$
& .09$^{+.01}_{-.02}$ & 1.37/94 & \nodata \\
& LH(d) & 52483 & 70124-02-06-00 & 0.4 & 1.67$^{+0.10}_{-0.11}$ &
1.14$^{+0.31}_{-0.27}$ & 1.56$^{+0.06}_{-0.06}$
& 0.076$^{+0.014}_{-0.012}$ & \nodata & \nodata & \nodata & 1.35/97 & \nodata \\
XTE~J1550-564 & P,VH & 51076 & 30191-01-02-00 & 2.0 &
4.18$^{+0.15}_{-0.16}$ & 2.25$^{+0.35}_{-0.24}$ &
2.92$^{+0.01}_{-0.01}$ & 234$^{+6}_{-6}$
& 5.9$^{+0.4}_{-0.5}$ & 1.72$^{+0.29}_{-0.17}$ & .68$^{+.23}_{-.22}$ & 1.95/97 & smedge at 8.5~keV\\
& HS & 51201 & 40401-01-19-00 & 2.0 & 1.137$^{+0.004}_{-0.004}$ &
3822$^{+85}_{-83}$ & 2.18$^{+0.02}_{-0.02}$ &
3.86$^{+0.24}_{-0.23}$& \nodata &
\nodata & \nodata & 1.81/101 & \nodata\\
& LH(r) & 51066 & 30188-06-01-03 & 2.0 & 0.78$^{+0.09}_{-0.02}$ &
1588$^{+111}_{-768}$ & 1.600$^{+0.008}_{-0.004}$ &
4.56$^{+0.14}_{-0.02}$ & 6.3$^{+0.6}_{-0.3}$ &
1.16$^{+0.07}_{-0.02}$ & .095$^{+.05}_{-.02}$ & 1.84/97 & PL
cutoff 35~keV\\
XTE~J1650-500 & P,VH & 52162 & 60113-01-05-01 & 0.6 &
1.08$^{+0.01}_{-0.01}$ & 176$^{+8}_{-11}$ & 1.52$^{+0.02}_{-0.02}$ &
1.55$^{+0.01}_{-0.01}$ & 6.3$^{+0.5}_{-0.5}$ &
1.42$^{+0.03}_{-0.06}$ & .057$^{+.002}_{-.003}$ & 1.79/94 & PL cutoff 97~keV\\
& HS & 52187 & 60113-01-25-00 & 0.6 & 0.664$^{+0.003}_{-0.003}$ &
8147$^{+201}_{-202}$ & 2.25$^{+0.02}_{-0.02}$ &
0.94$^{+0.06}_{-0.07}$ & 6.5(fixed) & 0.69$^{+0.11}_{-0.11}$
& .005$^{+.001}_{-.001}$ & 1.75/94 & smedge at 8~keV\\
&LH(r) & 52158 & 60113-01-01-00 & 0.6 & 1.66$^{+0.09}_{-0.17}$ &
21.1$^{+6.7}_{-2.6}$ & 1.33$^{+0.07}_{-0.06}$ &
0.85$^{+0.15}_{-0.12}$ & 6.53$^{+0.12}_{-0.17}$ &
0.80$^{+0.4}_{-0.3}$ & .012$^{+.010}_{-.004}$ & 1.52/93 & PL cutoff 84\\
GRO~J1655-40 & P,VH & 53508 & 91702-01-58-00 & 0.6 &
4.05$^{+0.19}_{-0.19}$ & 4.4$^{+1.1}_{-1.1}$
& 2.64$^{+0.03}_{-0.03}$ & 84$^{+2}_{-2}$ & 6.3$^{+0.4}_{-0.4}$
& 0.87$^{+0.15}_{-0.15}$ & .26$^{+.05}_{-.05}$ & 2.98/91 & smedge at 9~keV, PL cutoff\\
& HS & 53448 & 91702-01-07-02 & 0.6 & 1.324$^{+0.011}_{-0.011}$ &
1089$^{+50}_{-49}$ & 2.40$^{+0.08}_{-0.08}$ & 2.55$^{+0.74}_{-0.55}$
& \nodata & \nodata & \nodata & 0.81/94 & smedge at 7.1~keV\\
& LH(r)& 53440 & 91702-01-02-00G & 0.6 & 1.33$^{+0.02}_{-0.02}$ &
222$^{+11}_{-15}$ & 2.06$^{+0.05}_{-0.02}$ & 4.7$^{+0.3}_{-0.1}$
& \nodata & \nodata & \nodata & 1.59/96 & PL cutoff 144~keV\\
4U~1659-487 & P,HS & 52484 & 70110-01-33-00 & 0.4 &
0.888$^{+0.009}_{-0.009}$ & 3377$^{+210}_{-185}$ &
2.16$^{+0.08}_{-0.08}$ & 0.43$^{+0.09}_{-0.08}$
& 6.4(fixed) & 0.60$^{+0.15}_{-0.19}$ & .009$^{+.003}_{-.002}$ & 0.75/94 & \nodata \\
(GX 339-4) & LH(r) & 52398 & 70110-01-09-00 & 0.4 &
1.21$^{+0.12}_{-0.06}$ & 66$^{+11}_{-21}$ & 1.34$^{+0.04}_{-0.03}$ &
1.05$^{+0.07}_{-0.05}$ & 6.3$^{+1.1}_{-1.1}$ &
0.22$^{+0.09}_{-0.19}$&
 .045$^{+.006}_{-.011}$ & 1.29/94 & PL cutoff 42~keV\\
XTE~J1859+226 & P,HS & 51467 & 40124-01-12-00 & 0.2 &
1.14$^{+0.01}_{-0.01}$ & 841$^{+20}_{-81}$ &
2.54$^{+0.02}_{-0.02}$ & 15.1$^{+0.8}_{-0.7}$ & \nodata & \nodata & \nodata & 1.41/97 & \nodata \\
& LH(r) & 51462 & 40124-01-04-00 & 0.2 & 1.03$^{+0.14}_{-0.11}$ &
103$^{+82}_{-47}$ & 1.50$^{+0.05}_{-0.06}$ & 1.18$^{+0.13}_{-0.13}$
& 6.4(fixed) & 1.18$^{+0.18}_{-0.18}$ & .023$^{+.001}_{-.001}$ & 0.75/94 &PL cutoff 45~keV\\
GRS~1915+105 & P,HS & 51513 & 40703-01-40-01 & 8 &
2.79$^{+0.11}_{-0.11}$ & 14.1$^{+3.5}_{-2.7}$ &
3.33$^{+0.02}_{-0.02}$ & 328$^{+11}_{-11}$ & 6.4(fixed) &
1.27$^{+0.1}_{-0.1}$ & .356$^{+.04}_{-.04}$ & 1.25/95 & \nodata \\
& LH(r) & 51367 & 40403-01-09-00 & 8 & 0.69$^{+0.03}_{-0.01}$ &
5767$^{+1982}_{-1366}$ & 1.97$^{+0.02}_{-0.01}$ &
6.50$^{+0.23}_{-0.08}$ & 6.3$^{+1.0}_{-1.0}$ &
1.28$^{+0.06}_{-0.14}$
& .092$^{+.003}_{-.014}$ & 1.49/93 & PL cutoff 29~keV\\
\hline IGR~J00291+5934 & P,LH & 53342 & 90052-03-01-00 & 0.5 &
1.01$^{+0.06}_{-0.07}$ & 0.0012$^{+0.0003}_{-0.0003}$ &
1.38$^{+0.07}_{-0.08}$ & 0.085$^{+0.012}_{-0.012}$ & \nodata &
\nodata & \nodata & 1.07/96 & PL cutoff 79~keV \\
EXO~0748-676 & HS & 53483 & 91035-01-01-05 & 0.1 &
1.19$^{+0.05}_{-0.05}$ & 0.0073$^{+0.0006}_{-0.0005}$
& 3.1$^{+0.3}_{-0.2}$ & 0.7$^{+0.3}_{-0.2}$ & 6.4(fixed)& 4.59$^{+0.3}_{-0.3}$
& .020$^{+.002}_{-.002}$ & 0.87/95 & \nodata\\
& LH(d) & 53494 & 91035-01-04-13 & 0.1 & 1.60$^{+0.23}_{-0.15}$ &
0.0006$^{+0.0002}_{-0.0002}$ & 1.36$^{+0.13}_{-0.15}$ &
0.017$^{+0.007}_{-0.005}$ & \nodata & \nodata & \nodata & 0.94/98 & \nodata\\
XTE~J0929-314 & P,LH & 52403 & 70096-03-02-00 & 0.12 &
0.72$^{+0.06}_{-0.07}$ & 0.0009$^{+0.0001}_{-0.0001}$ &
1.79$^{+0.04}_{-0.04}$ & 0.084$^{+0.008}_{-0.008}$
& \nodata & \nodata & \nodata & 1.04/97 & \nodata\\
4U~1608-52 & P,HS & 50850 & 30062-03-01-02 & 1.8 & 2.15$^{+0.02}_{-0.02}$
& 0.25$^{+0.01}_{-0.01}$ & 3.76$^{+0.02}_{-0.02}$ & 78$^{+3}_{-3}$
& 6.4(fixed) & 1.87$^{+0.04}_{-0.04}$ & .50$^{+.04}_{-.04}$& 2.37/96
& smedge at 9~keV\\
&LH(d) & 50903 & 30062-01-01-04 & 1.8 & 0.73$^{+0.06}_{-0.05}$ &
0.0022$^{+0.0003}_{-0.0003}$ & 1.97$^{+0.02}_{-0.02}$ &
0.303$^{+0.015}_{-0.015}$ & 6.4(fixed) &
1.10$^{+0.15}_{-0.15}$ & .0024$^{+.0005}_{-.0005}$ & 0.90/99 & \nodata\\
MXB~1659-298 & P,HS & 51276 & 40050-04-05-01 & 0.2 &
2.15$^{+0.04}_{-0.04}$ & 0.0048$^{+0.0001}_{-0.0001}$ &
2.76$^{+0.03}_{-0.03}$ & 0.98$^{+0.04}_{-0.04}$ &
6.3$^{+0.5}_{-0.5}$ & 0.97$^{+0.13}_{-0.13}$
& .0033$^{+.0009}_{-.0009}$ & 2.58/94 & \nodata\\
XTE~J1710-281 & P,HS & 51359 & 40135-01-43-00 & 0.24 &
1.65$^{+0.11}_{-0.11}$ & 0.00046$^{+0.00005}_{-0.00005}$ &
2.52$^{+0.08}_{-0.07}$ & 0.098$^{+0.013}_{-0.012}$ & \nodata &
\nodata & \nodata & 0.54/97 &\nodata\\
1A~1744-361 & P,HS & 52965 & 80431-01-03-02 & 0.3 &
2.08$^{+0.03}_{-0.03}$ & 0.030$^{+0.001}_{-0.001}$ &
2.9$^{+0.08}_{-0.07}$& 2.2$^{+0.2}_{-0.2}$
& \nodata & \nodata & \nodata & 1.44/94 & smedge at 6.6~keV\\
& LH(d) & 53103 & 90058-04-01-00 & 0.3 & 1.60$^{+0.12}_{-0.09}$ &
0.0011$^{+0.0002}_{-0.0002}$ & 1.76$^{+0.06}_{-0.07}$ &
0.064$^{+0.011}_{-0.010}$ & \nodata & \nodata & \nodata & 0.89/98 & \nodata\\
GRS~1747-312 & P,HS & 51448 & 40419-01-02-00 & 0.6 &
2.22$^{+0.04}_{-0.06}$ & 0.0113$^{+0.0005}_{-0.0006}$ &
2.58$^{+0.09}_{-0.08}$ & 0.43$^{+0.06}_{-0.05}$ & \nodata & \nodata
& \nodata & 0.89/94 & smedge at 6.5~keV \\
& LH(d)\tablenotemark{a} & 53258 & 80045-02-10-01 & 0.6 &
1.42$^{+0.20}_{-0.29}$ & 0.0002$^{+0.0001}_{-0.0001}$ &
2.08$^{+0.17}_{-0.25}$ &
0.022$^{+0.009}_{-0.010}$ & \nodata & \nodata & \nodata & 0.94/98 & \nodata \\
XTE~J1751-305 & P,LH & 52369 & 70131-01-01-00 & 0.6 &
1.38$^{+0.08}_{-0.05}$ & 0.0020$^{+0.0003}_{-0.0004}$ &
1.45$^{+0.06}_{-0.05}$ & 0.164$^{+0.015}_{-0.022}$ & \nodata &
\nodata
& \nodata & 0.85/96 & PL cutoff 38~keV\\
XTE~J1807-294 & P,LH & 52697 & 70134-09-02-00 & 0.3 &
1.1$^{+0.3}_{-0.1}$ & 0.0004$^{+0.0002}_{-0.0002}$ &
1.72$^{+0.08}_{-0.06}$ & 0.16$^{+0.08}_{-0.02}$ & \nodata & \nodata
& \nodata & 0.90/96 & PL cutoff 47~keV\\
SAX~J1808.4-3658 & P,LH & 52563 & 70080-01-01-01 & 0.12 &
0.79$^{+0.03}_{-0.03}$ & 0.0085$^{+0.0006}_{-0.0006}$ &
1.51$^{+0.04}_{-0.05}$ & 0.27$^{+0.03}_{-0.02}$ & 6.4(fixed) &
1.14$^{+0.13}_{-0.12}$ & .0073$^{+0.0012}_{-0.0012}$ & 0.98/94 & PL
cutoff 37~keV\\
XTE~J1814-338 & P,LH & 52818 & 80418-01-04-01 & 0.15 &
1.15$^{+0.09}_{-0.09}$ & 0.0004$^{+0.0002}_{-0.0002}$ &
1.49$^{+0.14}_{-0.09}$ & 0.052$^{+0.008}_{-0.012}$ & \nodata &
\nodata & \nodata & 0.86/96 & PL cutoff 54~keV\\
GS~1826-238 & P,LH & 52484 & 70044-01-01-01 & 0.17 &
1.34$^{+0.17}_{-0.09}$ & 0.0014$^{+0.0004}_{-0.0005}$ &
1.29$^{+0.08}_{-0.06}$ & 0.14$^{+0.02}_{-0.01}$
& \nodata & \nodata & \nodata & 1.07/96 & PL cutoff 35~keV\\
HETE~J1900.1-2455 & P,LH & 53553 & 91015-01-05-00 & 0.10 &
1.11$^{+0.03}_{-0.03}$ & 0.0026$^{+0.0004}_{-0.0003}$ &
1.41$^{+0.12}_{-0.15}$ & 0.075$^{+0.018}_{-0.018}$ & \nodata &
\nodata & \nodata & 1.31/97 & PL cutoff 33~keV \\
4U~1908+005 & P,HS & 51834 & 50049-02-04-00 & 0.3 &
1.97$^{+0.02}_{-0.02}$ & 0.154$^{+0.004}_{-0.004}$ &
3.89$^{+0.07}_{-0.07}$ & 49$^{+4}_{-5}$ & 6.4(fixed) &
1.62$^{+0.11}_{-0.093}$
& .22$^{+.03}_{-0.03}$ & 1.90/92 & smedge at 6~keV\\
(Aql~X-1) & LH(r) & 51820 &  50049-01-04-03 & 0.3 &
1.39$^{+0.07}_{-0.08}$ & 0.010$^{+0.004}_{-0.004}$ &
1.29$^{+0.08}_{-0.08}$ & 0.68$^{+0.12}_{-0.11}$ & 6.4(fixed) &
1.16$^{+0.23}_{-0.27}$ & .016$^{+.006}_{-.006}$ & 0.83/94 & PL
cutoff 23~keV\\
XTE~J2123-058 & P,HS & 50991 & 30511-01-01-00 & 0.06 &
1.91$^{+0.04}_{-0.04}$ & 0.0109$^{+0.0004}_{-0.0004}$ &
2.51$^{+0.05}_{-0.05}$ & 0.99$^{+0.09}_{-0.08}$
& \nodata & \nodata & \nodata & 1.37/101 & \nodata\\
& LH(d) & 51040 & 30511-01-06-00 & 0.06 & 0.88$^{+0.06}_{-0.07}$ &
0.00015$^{+0.0.00004}_{-0.00003}$ & 1.72$^{+0.01}_{-0.02}$ &
0.015$^{+0.001}_{-0.001}$ & \nodata & \nodata & \nodata & 0.82/101 &
\nodata\\

\enddata \tablecomments{`Source'--the source name. The
sources above and under the horizontal line are BH systems and NS
systems respectively. `State'--the location and state of the
observation. The letter `P' indicates the observation corresponds to
the outburst peak, of which the spectral state is one of the
followings: the very high state (VH), the high/soft state (HS) and
the low/haed state (LH). The observation without the letter `P' has
the maximal flux of the corresponding state. The letter `r' or `d'
in the parentheses indicates the LH state is during the rising or
decaying phase of an outburst. `MJD'--the MJD of the observation.
`ObsID'--the observation ID of the {\it RXTE} pointed observation.
`$N_{\rm{H}}$'--the hydrogen column density, in units of
10$^{-22}$~cm$^{-2}$. `$T_{\rm{B}}$'--the temperature at inner disk
radius in ``diskbb'' model for BH systems, or the temperature of
blackbody in ``bbody'' model for NS systems, in the units of keV.
`$N_{\rm{B}}$'--the normalization of ``diskbb'' model for BH systems
or ``bbody'' model for NS systems, see the Xspec manual for details.
`$\Gamma_{\rm{PL}}$'--the photon index of powerlaw.
`$N_{\rm{PL}}$'--the normalization of powerlaw,
photons~keV$^{-1}$~cm$^{-2}$~s$^{-1}$ at 1~keV.
`$E_{\rm{Fe~line}}$'--the energy of Fe line, in the units of keV.
`$\sigma_{\rm{Fe~line}}$'--the width of Fe line, in the units of keV.
`$N_{\rm{Fe~line}}$'--normalization of Fe line, in the units of
photons~cm$^{-2}$~s$^{-1}$. `$\chi_{\nu}^2$/dof'--reduced
chi-square/degrees of freedom. `additional details'--additional
component of the spectrum. The error of each fitted parameter
corresponds to the 90\% confidence region.} \tablenotetext{a}{The
decay of a different outburst.}

\end{deluxetable}
\clearpage

\begin{deluxetable}{cccccccc}
\tablecaption{Parameters and Fluxes of transient LMXBs Plotted in
Figure~2--4 \label{tbl3}} \tablewidth{0pt}
\setlength{\tabcolsep}{0.04in}
\renewcommand\arraystretch{1.5} \tabletypesize{\scriptsize}
\tablehead{ \colhead{Mark} & \colhead{Sources} &
\colhead{$P\rm{_{orb}}$ (hr)} &
\colhead{$M$(M$_{\odot}$)\tablenotemark{b}} & \colhead{$D$(kpc)} &
\colhead{$F_{\rm{peak}}$} & \colhead{$F_{\rm{HS,max}}$} &
\colhead{$F_{\rm{LH,max}}$}}

\startdata

1 & XTE~J1118+480 & 4.08 & 8.53$\pm$0.06/6.5-7.2[M] & 1.8$\pm$0.5[M] & 4.12$^{+0.11}_{-0.09}$ & \nodata & 4.12$^{+0.11}_{-0.09}$\\
2 & GS~1354-64 & 61.07 & $\ge$5.75 & $\ge$27 & 5.60$^{+0.05}_{-0.12}$ & \nodata & 5.60$^{+0.05}_{-0.12}$\\
3 & 4U~1543-47 & 26.8 & 8.5-10.4 & 7.5$\pm$0.5[M] & 45.76$^{+0.17}_{-0.17}$ & 45.76$^{+0.17}_{-0.17}$ & 2.24$^{+0.05}_{-0.08}$\\
4 & XTE~1550-564 & 37.25 & 10.5$\pm$1.0 & 5.3$\pm$2.3[M] & 170.9$^{+2.1}_{-2.2}$ & 58.18$^{+0.16}_{-0.17}$ & 42.0$^{+0.6}_{-0.6}$\\
5 & XTE~J1650-500 & 7.63 & 2.73-7.3\tablenotemark{a} & 2-6 & 28.62$^{+0.20}_{-0.20}$ & 5.86$^{+0.04}_{-0.04}$ & 27.6$^{+0.3}_{-0.4}$\\
6 & GRO~J1655-40 & 62.88 & 7.02$\pm$0.22/6.0-6.6[M] & 3.2$\pm$0.2[M] & 112.9$^{+1.5}_{-1.3}$ & 32.1$^{+0.3}_{-0.3}$ & 26.50$^{+0.08}_{-0.10}$\\
7 & 4U~1659-487 & 42.14 & $\ge$5.8 & $>$6 & 10.99$^{+0.09}_{-0.07}$ & 10.99$^{+0.09}_{-0.07}$ & 23.4$^{+0.5}_{-0.5}$\\
8 & XTE~J1859+226 & 9.16 & 7.6-12.0[M] & 7.6/11[M] & 31.51$^{+0.11}_{-0.12}$ & 31.51$^{+0.11}_{-0.12}$& 15.66$^{+0.20}_{-0.24}$\\
9 & GRS~1915+105 & 804 & 14$\pm$4 & 11.2-12.5 & 107.4$^{+0.4}_{-0.4}$ & 107.4$^{+0.4}_{-0.4}$ & 24.2$^{+0.4}_{-0.4}$\\
\hline

a & IGR~J00291+5934 & 2.46 & \nodata & 2.6-3.6 & 2.35$^{+0.06}_{-0.16}$ & \nodata & 2.35$^{+0.06}_{-0.16}$\\
b & EXO~0748-676 & 3.82 & \nodata & 5.9-7.7 & 0.936$^{+0.023}_{-0.022}$ & 0.936$^{+0.023}_{-0.022}$ & 0.97$^{+0.0.09}_{-0.13}$\\
c & XTE~J0929-314 & 0.73 & \nodata & 10$\pm$5 & 1.15$^{+0.05}_{-0.05}$ & \nodata & 1.15$^{+0.05}_{-0.05}$\\
d & 4U~1608-52 & 12.89 & \nodata & 4.1/3.6 & 29.0$^{+1.6}_{-1.6}$ & 29.0$^{+1.6}_{-1.6}$ & 2.150$^{+0.031}_{-0.033}$\\
e & MXB~1659-298 & 7.11 & \nodata & 10 & 1.215$^{+0.018}_{-0.014}$ & 1.215$^{+0.018}_{-0.014}$ &\nodata\\
f & XTE~J1710-281 & 3.28 & \nodata & 12-16 & 0.179$^{+0.024}_{-0.019}$ & 0.179$^{+0.024}_{-0.019}$ &\nodata\\
g & 1A~1744-361 & 1.62 & \nodata & $<$9 & 3.29$^{+0.22}_{-0.22}$ & 3.29$^{+0.22}_{-0.22}$ & 0.98$^{+0.07}_{-0.07}$\\
h & GRS~1747-312 & 12.36 & \nodata & 9.5 & 1.294$^{+0.022}_{-0.020}$ & 1.294$^{+0.022}_{-0.020}$ & 0.17$^{+0.05}_{-0.06}$\\
i & XTE~J1751-305 & 0.71 & \nodata & 8.5/$>$7 & 2.25$^{+0.02}_{-0.03}$ & \nodata & $2.25^{+0.02}_{-0.03}$\\
j & XTE~J1807-294 & 0.668 & \nodata & 8\tablenotemark{c} & 1.20$^{+0.02}_{-0.04}$ & \nodata & 1.20$^{+0.02}_{-0.04}$\\
k & SAX~J1808.4-3658 & 2.014167 & $<$2.27 & 2.5/3.4-3.6 & 3.44$^{+0.19}_{-0.22}$ & \nodata & 3.44$^{+0.19}_{-0.22}$\\
l & XTE~J1814-338 & 4.27462 & \nodata & 8.0$\pm$1.6 & 0.78$^{+0.02}_{-0.04}$ & \nodata & 0.78$^{+0.02}_{-0.04}$\\
m & GS~1826-238 & 2.088 & \nodata & 4-8 & 2.88$^{+0.04}_{-0.05}$ & \nodata & 2.88$^{+0.04}_{-0.05}$\\
n & HETE~J1900.1-2455 & 1.39 & \nodata & 5 & 1.15$^{+0.03}_{-0.06}$ & \nodata & 1.15$^{+0.03}_{-0.06}$\\
o & 4U~1908+005 & 18.95 & \nodata & 5(4-6.5) & 16.2$^{+1.2}_{-1.2}$ & 16.2$^{+1.2}_{-1.2}$ & 9.98$^{+0.08}_{-0.11}$\\
p & XTE~J2123-058 & 5.96 & 1.5$\pm$0.3/1.04-1.56 & 8.5$\pm$2.5 & 2.26$^{+0.07}_{-0.06}$ & 2.26$^{+0.07}_{-0.06}$ & 0.191$^{+0.018}_{-0.020}$\\

\enddata

\tablecomments{The first column is the mark of a source (number for
BH and letter for NS), which is used in
Figure~\ref{peaklum}--\ref{lhlum}. The mass and distance data are
from LPH07 and references therein, except those marked with `M' in
the square brackets which are from MR06. `$F\rm{_{peak}}$',
`$F_{\rm{HS,max}}$', `$F_{\rm{HS,max}}$' are the unabsorbed fluxes
of outburst peak, the HS state maximum and the LH state maximum
respectively, in the units of 10$^{-9}~\rm{erg~cm^{-2}~s^{-1}}$.}
\tablenotetext{a}{Mass function f($M$)=2.73~M$_{\odot}$ and the BH
mass $<$ 7.3~M$_{\odot}$ (LPH07).} \tablenotetext{b}{The mass of NS
is assumed in the range of 1.4--2.2~M$_{\odot}$ if it is not known.}
\tablenotetext{c}{The distance is unknown, but usually assumed as 8
kpc as the source might be close to the Galactic center
\citep[e.g.][]{Cam03,Fal05}.}
\end{deluxetable}
\clearpage


\clearpage

\begin{thebibliography}{}
\bibitem[Altamirano et al.(2008)]{Alt08} Altamirano, D.,
Casella, P., Patruno, A., Wijnands, R., \& van der Klis, M.\ 2008,
\apjl, 674, L45
\bibitem[Barziv et al.(2001)]{Bar01} Barziv, O., Kaper, L., Van Kerkwijk, M.~H.,
Telting, J.~H., \& Van Paradijs, J.\ 2001, \aap, 377, 925
\bibitem[Belloni et al.(2005)]{Bel05} Belloni, T., Homan, J.,
Casella, P., van der Klis, M., Nespoli, E., Lewin, W.~H.~G., Miller,
J.~M., \& M{\'e}ndez, M.\ 2005, \aap, 440, 207
\bibitem[Bhattacharyya et al.(2006)]{Bha06} Bhattacharyya,
S., Strohmayer, T.~E., Markwardt, C.~B., \& Swank, J.~H.\ 2006,
\apjl, 639, L31
\bibitem[Brocksopp et al.(2006)]{Bro06} Brocksopp, C., et
al.\ 2006, \mnras, 365, 1203
\bibitem[Burderi et al.(2006)]{Bur06} Burderi, L., et al.\
2006, Chinese Journal of Astronomy and Astrophysics Supplement, 6,
192
\bibitem[Campana et al.(2003)]{Cam03} Campana, S., Ravasio,
M., Israel, G.~L., Mangano, V., \& Belloni, T.\ 2003, \apjl, 594,
L39
\bibitem[Cannizzo(1996)]{Can96} Cannizzo, J.~K.\ 1996, \apjl,
466, L31
\bibitem[Casares et al.(2004)]{Cas04} Casares, J., Zurita,
C., Shahbaz, T., Charles, P.~A., \& Fender, R.~P.\ 2004, \apjl, 613,
L133
\bibitem[Casella et al.(2008)]{Cas08} Casella, P.,
Altamirano, D., Patruno, A., Wijnands, R., \& van der Klis, M.\
2008, \apjl, 674, L41
\bibitem[Chen et al.(1997)]{CSL97} Chen, W., Shrader, C.~R., \& Livio, M.\ (CSL97)
1997, \apj, 491, 312
\bibitem[Clark et al.(2002)]{Cla02} Clark, J.~S., Goodwin, S.~P.,
Crowther, P.~A., Kaper, L., Fairbairn, M., Langer, N., \& Brocksopp,
C.\ 2002, \aap, 392, 909
\bibitem[Cornelisse et al.(2007)]{Cor07} Cornelisse, R.,
Casares, J., Steeghs, D., Barnes, A.~D., Charles, P.~A., Hynes,
R.~I., \& O'Brien, K.\ 2007, \mnras, 375, 1463
\bibitem[Dickey \& Lockman(1990)]{DL90} Dickey, J.~M., \&
Lockman, F.~J.\ 1990, \araa, 28, 215
\bibitem[Done et al.(2007)]{Don07} Done, C., Gierli{\'n}ski, M., \&
Kubota, A.\ 2007, \aapr, 15, 1
\bibitem[Eggleton(1983)]{Egg83} Eggleton, P.~P.\ 1983, \apj,
268, 368 
\bibitem[Esin et al.(2001)]{Esin01}Esin, A.~A., McClintock,
J.~E., Drake, J.~J., Garcia, M.~R., Haswell, C.~A., Hynes, R.~I., \&
Muno, M.~P.\ 2001, \apj, 555, 483
\bibitem[Esin et al.(1997)]{Esin97}Esin, A. A., McClintock, J. E., \& Narayan, R. 1997, \apj, 489, 865
\bibitem[Falanga et al.(2005)]{Fal05} Falanga, M., et al.\
2005, \aap, 436, 647
\bibitem[Fender et al.(2004a)]{FBG04} Fender, R.~P., Belloni,
T.~M., \& Gallo, E.\ 2004a, \mnras, 355, 1105
\bibitem[Fender et al.(2004b)]{Fen04} Fender, R., De Bruyn,
G., Pooley, G., \& Stappers, B.\ 2004b, The Astronomer's Telegram,
361, 1
\bibitem[Frank et al.(1992)]{Fra92} Frank, J., King, A., \& Raine, D. 1992, Accretion Power in
Astrophysics (Cambridge: Cambridge Univ. Press)
\bibitem[Garcia et al.(2000)]{Gar00} Garcia, M., Brown, W., Pahre, M., McClintock, J.,
Callanan, P., \& Garnavich, P.\ 2000, \iaucirc, 7392, 2
\bibitem[Garcia et al.(2001)]{Gar01} Garcia, M.~R.,
McClintock, J.~E., Narayan, R., Callanan, P., Barret, D., \& Murray,
S.~S.\ 2001, \apjl, 553, L47
\bibitem[Garcia et al.(2003)]{Gar03} Garcia, M.~R., Miller,
J.~M., McClintock, J.~E., King, A.~R., \& Orosz, J.\ 2003, \apj,
591, 388
\bibitem[Giles et al.(1996)]{Gil96} Giles, A.~B., Swank,
J.~H., Jahoda, K., Zhang, W., Strohmayer, T., Stark, M.~J., \&
Morgan, E.~H.\ 1996, \apjl, 469, L25
\bibitem[Heiselberg \& Pandharipande(2000)]{HP00} Heiselberg, H., \& Pandharipande, V.\
2000, Annual Review of Nuclear and Particle Science, 50, 481
\bibitem[Homan et al.(2005)]{Hom05} Homan, J., Buxton, M.,
Markoff, S., Bailyn, C.~D., Nespoli, E., \& Belloni, T.\ 2005, \apj,
624, 295
\bibitem[Homan et al.(2001)]{Hom01} Homan, J., Wijnands, R.,
van der Klis, M., Belloni, T., van Paradijs, J., Klein-Wolt, M.,
Fender, R., \& M{\'e}ndez, M.\ 2001, \apjs, 132, 377
\bibitem[Johnston et al.(1999)]{Joh99} Johnston, H.~M.,
Fender, R., \& Wu, K.\ 1999, \mnras, 308, 415
\bibitem[Jonker et al.(2007a)]{Jon07} Jonker, P.~G., Nelemans,
G., \& Bassa, C.~G.\ 2007a, \mnras, 374, 999
\bibitem[Jonker et al.(2007b)]{Jon07b} Jonker, P.~G., Steeghs,
D., Chakrabarty, D., \& Juett, A.~M.\ 2007b, \apjl, 665, L147
\bibitem[Kalberla et al.(2005)]{Kal05} Kalberla, P.~M.~W.,
Burton, W.~B., Hartmann, D., Arnal, E.~M., Bajaja, E., Morras, R.,
\& {P\"o}ppel, W.~G.~L.\ 2005, \aap, 440, 775
\bibitem[King(1988)]{Kin88} King, A.~R.\ 1988, \qjras, 29, 1
\bibitem[King et al.(1996)]{Kin96} King, A.~R., Kolb, U., \&
Burderi, L.\ 1996, \apjl, 464, L127
\bibitem[King \& Ritter(1998)]{KR98} King, A.~R., \& Ritter, H.\
1998, \mnras, 293, L42
\bibitem[Kouveliotou et al.(1996)]{Kou96} Kouveliotou, C.,
van Paradijs, J., Fishman, G.~J., Briggs, M.~S., Kommers, J.,
Harmon, B.~A., Meegan, C.~A., \& Lewin, W.~H.~G.\ 1996, \nat, 379,
799
\bibitem[Krimm et al.(2007)]{Kri07} Krimm, H.~A., et al.\
2007, \apjl, 668, L147
\bibitem[Lasota(2001)]{Las01} Lasota, J.-P.\ 2001, New
Astronomy Review, 45, 449
\bibitem[Liu et al.(2007)]{Liu07} Liu, Q.~Z., van Paradijs,
J., \& van den Heuvel, E.~P.~J.\ 2007, \aap, 469, 807 (LPH07)
\bibitem[Lubow(1991)]{Lub91} Lubow, S.~H.\ 1991, \apj, 381,
268
\bibitem[Maccarone \& Coppi(2003)]{MC03} Maccarone, T.~J., \& Coppi, P.~S.\ 2003,
\mnras, 338, 189
\bibitem[Malzac et al.(2004)]{Mal04} Malzac, J., Merloni, A.,
\& Fabian, A.~C.\ 2004, \mnras, 351, 253
\bibitem[Markoff et al.(2001)]{Mar01} Markoff, S., Falcke, H., \& Fender, R.\ 2001,
\aap, 372, L25
\bibitem[Markwardt et al.(2002)]{Mar02} Markwardt, C.~B.,
Swank, J.~H., Strohmayer, T.~E., in 't Zand, J.~J.~M.,
\& Marshall, F.~E.\ 2002, \apjl, 575, L21
\bibitem[Masetti et al.(1996)]{Mas96} Masetti, N., Bianchini,
A., Bonibaker, J., della Valle, M., \& Vio, R.\ 1996, \aap, 314, 123
\bibitem[McClintock \& Remillard(2006)]{MR06} McClintock,
J.~E., \& Remillard, R.~A.\ 2006, Compact stellar X-ray sources, 157
(MR06)
\bibitem[Menou et al.(1999)]{Men99} Menou, K., Esin, A.~A.,
Narayan, R., Garcia, M.~R., Lasota, J.-P., \& McClintock, J.~E.\
1999, \apj, 520, 276
\bibitem[Meyer-Hofmeister \& Meyer(2000)]{MM00}
Meyer-Hofmeister, E., \& Meyer, F.\ 2000, \aap, 355, 1073
\bibitem[Meyer-Hofmeister(2004)]{Mey04} Meyer-Hofmeister, E.\
2004, \aap, 423, 321
\bibitem[Miller et al.(2002)]{Mil02} Miller, J.~M., et al.\
2002, \apjl, 570, L69
\bibitem[Miyamoto et al.(1995)]{Miy95}
Miyamoto, S., Kitamoto, S., Hayashida, K., Egoshi, W. 1995, \apj,
442, L13
\bibitem[Muno et al.(2005)]{Mun05} Muno, M.~P., Pfahl, E.,
Baganoff, F.~K., Brandt, W.~N., Ghez, A., Lu, J., \& Morris, M.~R.\
2005, \apjl, 622, L113
\bibitem[Nowak(1995)]{Now95} Nowak, M.~A.\ 1995, \pasp, 107,
1207
\bibitem[Orosz et al.(2004)]{Oro04} Orosz,
J.~A., McClintock, J.~E., Remillard, R.~A., \& Corbel, S.\ 2004,
\apj, 616, 376
\bibitem[Paczynski(1977)]{Pac77} Paczynski, B.\ 1977, \apj,
216, 822
\bibitem[Park et al.(2004)]{Par04} Park, S.~Q., et al.\ 2004,
\apj, 610, 378
\bibitem[Parmar et al.(1986)]{Par86} Parmar, A.~N., White,
N.~E., Giommi, P., \& Gottwald, M.\ 1986, \apj, 308, 199
\bibitem[Portegies Zwart et al.(2004)]{PZ04} Portegies
Zwart, S.~F., Dewi, J., \& Maccarone, T.\ 2004, \mnras, 355, 413
\bibitem[Remillard et al.(2005)]{Rem05} Remillard, R.,
Garcia, M., Torres, M.~A.~P., \& Steeghs, D.\ 2005, The Astronomer's
Telegram, 384, 1
\bibitem[Shahbaz et al.(1998)]{Sha98} Shahbaz, T., Charles,
P.~A., \& King, A.~R.\ 1998, \mnras, 301, 382
\bibitem[Shahbaz et al.(1997)]{Sha97} Shahbaz, T., Naylor,
T., \& Charles, P.~A.\ 1997, \mnras, 285, 607
\bibitem[Shaposhnikov et al.(2007)]{Sha07} Shaposhnikov, N.,
Swank, J., Shrader, C.~R., Rupen, M., Beckmann, V., Markwardt,
C.~B., \& Smith, D.~A.\ 2007, \apj, 655, 434
\bibitem[Smith et al.(2002)]{SHS02}
Smith, D. M., Heindl, W. A., Swank, J. 2002, \apj, 569, 362
\bibitem[Sobczak et al.(2000)]{Sob00} Sobczak, G.~J.,
McClintock, J.~E., Remillard, R.~A., Cui, W., Levine, A.~M., Morgan,
E.~H., Orosz, J.~A., \& Bailyn, C.~D.\ 2000, \apj, 544, 993
\bibitem[Strickman et al.(1996)]{Str96} Strickman, M.~S., et al.\
1996, \apjl, 464, L131
\bibitem[Tanaka \& Lewin(1995)]{TL95} Tanaka, Y., \& Lewin,
W.~H.~G.\ 1995, X-ray binaries, p.~126 - 174, 126
\bibitem[Tananbaum et al.(1972)]{Tan72} Tananbaum, H.,
Gursky, H., Kellogg, E., Giacconi, R., \& Jones, C.\ 1972, \apjl,
177, L5
\bibitem[van der Klis(1995)]{vdK95} van der Klis, M.\ 1995,
X-ray binaries, p.~252 - 307, 252
\bibitem[van Paradijs(1996)]{vP96} van Paradijs, J.\ 1996,
\apjl, 464, L139
\bibitem[van Paradijs \& McClintock(1995)]{vM95} van
Paradijs, J., \& McClintock, J.~E.\ 1995, X-ray binaries, p.~58 -
125, 58
\bibitem[Whitehurst(1988)]{Whi88} Whitehurst, R.\ 1988,
\mnras, 232, 35
\bibitem[Wu et al.(2009)]{Wu09} Wu, Y. X., Yu, W., Yan, Z., Sun, L.,
Li, T. P., submitted to \aap
\bibitem[Yu \& Dolence(2007)]{YD07} Yu, W., \& Dolence, J.\
2007, \apj, 667, 1043
\bibitem[Yu et al.(2007)]{YFK07} Yu, W., Lamb, F.~K., Fender,
R., \& van der Klis, M.\ 2007, \apj, 663, 1309
\bibitem[Yu, van der Klis \& Fender (2004)]{YKF04}
Yu, W., van der Klis, M. \& Fender, R., 2004, \apj, 611, L121
\bibitem[Yu \& Yan(2009)]{YY09} Yu, W., \& Yan, Z.\ 2009, \apj,
701, 1940
\bibitem[Zhang et al.(2006)]{Zhang06} Zhang, F., Qu, J. L., C. M. Zhang, et al. 2006, ApJ, 646, 1116
\end{thebibliography}
\end{document}